# Is There A Cosmological Constant?

CHRISTOPHER S. KOCHANEK

Harvard–Smithsonian Center for Astrophysics, 60 Garden St., Cambridge, MA 02138

I: ckochanek@cfa.harvard.edu

## ABSTRACT

We present limits on the cosmological constant from the statistics of gravitational lenses using newly completed quasar surveys, new lens data, and a range of lens models. The formal limit is $\lambda_0 < 0.66$ at 95% confidence in flat cosmologies, including the statistical uncertainties in the number of lenses, galaxies, quasars, and the parameters relating galaxy luminosities to dynamical variables. The limit holds for either softened isothermal or de Vaucouleurs models of the lens galaxies, suggesting that the radial mass distribution in the lens galaxy is not a significant systematic uncertainty. The cosmological limits are unaffected by adding a small core radius to the isothermal lens models or by the luminosity of the lens galaxy. Inconsistent models of the effects of a core radius make significant errors in the cosmological constraints. Extinction in E/S0 galaxies can significantly reduce the limits on the cosmological constant, but changing the expected number of lenses by a factor of two requires a minimum of 100 times more dust than is observed locally in E/S0 galaxies. Rapid evolution of the lenses is the most promising means of evading these limits. However, a viable model must not only reduce the expected number of lenses but do so without significantly altering the average image separations, magnitudes, redshifts, and the relative properties of optically and radio selected lenses.

*Subject headings:* cosmology: observations – galaxies: distances and redshifts – gravitational lensing – galaxies: structure – galaxies: kinematics and dynamics

## 1   INTRODUCTION

Ostriker & Steinhardt (1995) invoke a "cosmic concordance" of models and data for flat cosmologies with a cosmological constant of $\lambda_0 \simeq 0.65 \pm 0.1$. This value is already in disagreement with the most recent analysis of the statistics of gravitational lenses, which found that $\lambda_0 \lesssim 0.5$ at 90% confidence (Kochanek 1995). The statistical models were, however, limited in scope and included neither all the possible uncertainties nor recently discovered gravitational lenses or completed surveys, although the study treated some systematic uncertainties such as the mass distribution in E/S0 galaxies. The current concordance is already a retreat from earlier invocations of $\lambda_0 \sim 0.8$ (e.g. Efstathiou, Sutherland, & Maddox 1990) that were ruled out in the first full analyses of the observed statistics of gravitational lenses (Kochanek 1993, Maoz & Rix 1993). The dependence of the number of lenses on the cosmological constant is steeper at $\lambda_0 \sim 0.8$ than $\lambda_0 \sim 0.6$, so it is possible that a careful treatment of the uncertainties will broaden the error bars sufficiently to permit the concordance.

Turner (1990) first pointed out that cosmological models with a large cosmological constant produce large numbers of gravitational lenses. A flat cosmology with $\lambda_0 = 1$ has roughly ten times as many gravitational lenses as a flat cosmology with $\lambda_0 = 0$. A flat model with $\lambda_0 = 0.5$ and an open model with $\lambda_0 = 0$ and $\Omega_0 = 0$ both predict about twice as many lenses as a flat model with $\lambda_0 = 0$ and $\Omega_0 = 1$. We only treat flat models in this paper, but results for open cosmologies with $\lambda_0 = 0$ can be approximated by mapping the range $0.5 < \Omega_0 < 1$ with $\Omega_0 + \lambda_0 = 1$ into the range $0 < \Omega_0 < 1$ with $\lambda_0 = 0$. The full dependence of lens statistics on both $\lambda_0$ and $\Omega_0$ is given in Carroll, Press, & Turner (1992) or Kochanek (1993).

The first limits on the cosmological constant were derived from surveys for lensed quasars by Crampton et al. (1992), Yee et al. (1993), the Snapshot Survey (Bahcall et al. 1992, Maoz et al. 1992, 1993ab), and the ESO/Liège Survey (Surdej et al. 1993). These surveys examined 648 quasars and found five gravitational lenses. At the same time the MG (Burke et al. 1992) and JVAS (Patnaik et al. 1992) radio surveys found eight lenses. There are two in depth analyses of these data sets. Kochanek (1993) analyzed all the optical surveys and the radio lenses, but focused on a single lens model, and Maoz & Rix (1993) analyzed only the Snapshot Survey data while exploring a wide range of mass models and observational scaling laws.

Kochanek (1993) used only the singular isothermal sphere (SIS) model for the lenses and focused on cosmological limits and dynamical normalizations. The upper limit on the cosmological constant was $\lambda_0 < 0.8$ at 90% confidence using systematic assumptions biased in favor of a high cosmological constant. The characteristic dark matter velocity dispersion of an $L_*$ galaxy was $\sigma_{DM*} \simeq 245 \pm 30$ km s$^{-1}$ at 90% confidence. This value was lower than predicted by Turner, Ostriker & Gott (1984), who estimated that the velocity dispersion of the dark matter should be $(3/2)^{1/2}$ larger than the central velocity dispersion of the stars ($\sigma_{DM*} \simeq 275$ km s$^{-1}$). Better dynamical models by Kochanek (1993) and Kochanek (1994) (see also Breimer & Sanders (1993) and Franx (1993)) demonstrated that the assumptions leading to the $(3/2)^{1/2}$ correction were incorrect; dynamically consistent models have $\sigma_{DM*} \simeq 225 \pm 10$ km s$^{-1}$, consistent with the estimates from fitting the observed lens separations.

Maoz & Rix (1993) emphasized the effects of the mass distribution in the E/S0 galaxies rather than the cosmological limits, although they found a limit of $\lambda_0 \lesssim 0.7$ at 95% confidence. When Maoz & Rix (1993) tried de Vaucouleurs (1948) models for the lens galaxies using the dynamically estimated mass-to-light ratio of $(M/L)_B \simeq (10 \pm 2)h$ where $H_0 = 100h$ km s$^{-1}$ Mpc$^{-1}$ (e.g. van der Marel 1991), they found that the models were incapable of producing the observed image separations. Kochanek (1995) later demonstrated that $(M/L)_B \simeq (22 \pm 3)h$ was required for de Vaucouleurs mass models to fit the lens separations. Maoz & Rix (1993) then added a softened isothermal halo, and models with a sufficiently large halo could produce the observed image separations. The halo models were not designed to be consistent with stellar dynamics (Breimer & Sanders 1993), but the total mass distribution of the successful models closely resembles the isothermal spheres used by Kochanek (1993).

The dynamical conclusions of the statistical models ($\sigma_{DM*} \simeq \sigma_c$ and $(M/L)_B \simeq 22h$) are supported by other observations. Detailed models of the radio ring lens MG 1654+1346 by Kochanek (1995) also found that a nearly singular isothermal model fit the data extremely well and required a velocity dispersion comparable to the estimated dispersion of the stars, while a de Vaucouleurs model fit the data less well and required a mass-to-light ratio of $(M/L)_B \simeq (20.4 \pm 2.8)h$. X-ray studies of ellipticals (e.g. Fabbiano 1989) and studies of disks and rings (e.g. Kent 1990; de Zeeuw 1991) also suggest that galaxies are embedded in dark matter halos, and stellar dynamical models are consistent with both constant mass-to-light ratios and dark matter (e.g. de Zeeuw & Franx 1991; van der Marel 1991; Saglia, Bertin, & Stiavelli 1992).

Since these analyses, the NOT survey (Jaunsen et al. 1995) and the FKS survey (Kochanek, Falco, & Schild 1995) have increased the survey sample to 864 quasars with $z > 1$ and added one more probable lens (LBQS 1009-0252, Hewett et al. 1994, Surdej et al. 1994) to the statistical sample. At the same time, the MG (see Patnaik 1994), JVAS (see Patnaik 1994), and CLASS (Jackson et al. (1995), Meyers et al. (1995)) surveys have reached a total of eleven radio lenses. While we cannot accurately compute the absolute probability a radio source is lensed, we can reliably estimate the relative probability of the image separations. The observed properties of the radio lenses are powerful constraints on statistical models of the quasar surveys (Kochanek 1993).

Several other systematic errors are treated in the literature. Galaxy evolution was treated by Mao (1991), Mao & Kochanek (1994) and Rix et al. (1994), who found that the conclusions of the no-evolution statistical models are correct unless galaxy evolution is still very dramatic at $z \lesssim 1$. Appealing to evolution to save models with a high cosmological constant is awkward, because one effect of a cosmological constant is to make galaxies form and evolve earlier than in a standard $\Omega_0 = 1$ CDM cosmology (see Carroll, Press, & Turner 1992). Moreover, most simple merger models preserved the expected number of lenses. Kochanek (1991) discussed the selection effects associated with quasar surveys, the brightness of the lens galaxy, and reddening. For lens surveys of bright quasars, we expect only reddening to be a serious concern. Some of these effects were reconsidered by Tomita (1995) and Fukugita & Peebles (1995).

Kochanek (1992) noted that the relative redshifts of lens galaxies and their sources are an independent method of determining the cosmological model from the statistical approaches used to analyze the quasar surveys. The test is weaker than a full analysis, but even it suggests that $\lambda_0 \lesssim 0.9$. Recently King (1994) (also Helbig & Kayser (1995)) strongly criticized this analysis for ignoring the measurability of the lens redshift and claim that the observable lens redshifts are insensitive to the cosmological model. Since the lens redshift test avoids some systematic errors in the standard statistical analyses (particularly problems associated with completeness, the number density of galaxies, and the average extinction) it is worth reevaluating the test in light of the critique.

The primary goal of this paper is to set firmer limits on the cosmological constant in flat cosmological models. In §2 we discuss the statistical model and its observational inputs. In §3 we discuss the effects of the uncertainties in the observational inputs on the cosmological limits using a softened isothermal sphere to model the lens galaxies. In §4 we reconsider the de Vaucouleurs lens models to see whether the cosmological limits are affected by changes in the radial structure of the lens galaxies. In §5 we continue our analysis of the softened isothermal models by examining the effects of finite core radii on both dynamical models of galaxies and lens statistics. In §6 we consider the effects of spiral galaxies, and in §7 we consider the effects of the lens galaxy luminosity and extinction. In §8 we reevaluate the lens redshift test, and in §9 we summarize the results.

## 2  Methodology, Data, & Models

In this section we outline the calculations and elaborate on improvements and changes from earlier discussions by Kochanek (1991, 1992, 1993) in the selection effects models, calculation techniques, statistics, number counts, and models for galaxies. We start by reviewing the numerical and statistical approaches. Next we discuss lens models and their parameters. Finally we discuss the number counts of galaxies and quasars.

### 2.1  Calculation of Likelihoods

If the two-dimensional lens potential is a function of radius $\phi(r)$, then the lens equation is simply
$$u = x - \frac{x}{r}\frac{\partial \phi}{\partial r} \tag{1}$$
where $u$ is the angular coordinate on the source plane and $x$ is the angular coordinate on the image plane. The two-dimensional potential $\phi(r)$ satisfies the Poisson equation in angular coordinates $\nabla^2 \phi = 2\Sigma/\Sigma_c$ where $\Sigma_c = c^2 D_{OS}^A D_{OL}^A / 4\pi G D_{LS}^A$ is the critical surface density per square angle. The angular diameter distances $D_{OS}^A$, $D_{OL}^A$, and $D_{LS}^A$ relate proper distances to angular sizes between the observer and the source, the observer and the lens, and the lens and the source respectively (see Schneider et al. 1992). $D_{OL}$, $D_{LS}$ and $D_{OS}$ denote the proper motion distances between the observer, lens, and source planes.

If the surface density declines monotonically, the lens has two critical radii on which the magnification diverges (see Schneider, Ehlers, & Falco 1992, Blandford & Narayan 1986). The inverse magnification is
$$M^{-1} = \left[1 - \frac{1}{r}\frac{\partial \phi}{\partial r}\right]\left[1 - \frac{\partial^2 \phi}{\partial r^2}\right], \tag{2}$$
where the two factors are the eigenvalues of the inverse magnification tensor. The outer, or tangential, critical radius $r_+$ is the solution of $r - \phi_{,r} = 0$, and the inner, or radial, critical radius $r_-$ is the solution of $1 - \phi_{,rr} = 0$. The caustics associated with the two critical lines are $u_+ = 0$ for the tangential critical line (a pseudo-caustic technically) and $u_-$ for the radial critical line. When two images are merging on the radial critical line, the third image is located on the same side of the lens as the source at radius $r_{out}$. Sources inside the radial caustic are triply imaged, and sources outside the radial caustic are singly imaged. If we consider sources lying along an axis with $u > 0$, then the outermost image 1 lies on the same side of the lens as the source between the tangential critical line and the radius of the last multiple image ($r_+ \leq x_1 \leq r_{out}$). The other two images are on the opposite side of the lens from the source, with image 2 lying between the radial and tangential critical lines ($-r_+ \leq x_2 \leq -r_-$), and image 3 lying between the origin and the radial critical line ($-r_- \leq x_3 \leq 0$). When the source lies on the tangential caustic, images 1 and 2 are infinitely magnified creating an Einstein ring, and when the source lies on the radial caustic image 2 and 3 are infinitely magnified. An extensive discussion of the asymptotic statistical properties of these lenses is given in Blandford & Kochanek (1987).

The comoving volume element at proper motion distance $D_{OL}$ is

$$dV = \frac{4\pi D_{OL}^2 dD_{OL}}{(1 + \Omega_k D_{OL}^2/r_H^2)^{1/2}} \qquad (3)$$

where the "curvature density" is $\Omega_K = 1 - \Omega_0 - \lambda_0$, the matter density is $\Omega_0$, and the cosmological constant is $\lambda_0$ (Carroll, Press, & Turner 1992). If the total magnification of the three images for a source at position $u$ is $M$, then the probability, $p(m, z_s)$, that a source with magnitude $m$ and redshift $z$ is lensed, assuming a survey selection function $S(u)$ and a configuration selection function $C(u)$, is

$$p(m, z_s) = 2\pi \int_0^\infty dL \frac{dn}{dL} \int_0^{z_s} dV \int_0^{u_-} \frac{udu}{4\pi} S(u) C(u) \frac{dN}{dm}(m + 2.5 \log M(u)) \left[\frac{dN}{dm}(m)\right]^{-1}. \qquad (4)$$

The comoving density of lenses labeled by luminosity $L$ is $dn/dL$, and the image positions, magnifications, and caustics must be computed for each luminosity and redshift. The location of the outer caustic ($u_-$) and the magnification are functions of the lens model, the redshifts, and the luminosity.

The selection function $S(u)$ is equal to one when the current pattern of lensed images is detectable given their magnitudes, separations, and the survey selection function, and zero when it is not detectable. We consider the detectability of images in pairs (e.g. Webster et al. 1988, Kochanek 1991), and an unresolved close pair (typically the 2-3 images) is merged into an average image before comparing it to the remaining image. The configuration selection function $C(u)$ selects lenses with particular characteristics such as the image separations or the visibility of the central image. To estimate the probability of a particular image separation we calculated the probability of finding a lens within $\pm 10\%$ of the observed separation.

The calculation is numerically simpler if we integrate over the image plane positions of image 2 rather than over the source positions. This helps stablize the integration because the Jacobian of the transformation from the image plane to the source plane introduces a factor $M_2^{-1}$ into the integral and suppresses the near divergences of the integrand near the caustics. We also need only solve the lens equation (1) for the positions of the 1 and 3 images. The lens probability with the change of integration variables is

$$p(m, z_s) = \frac{1}{2} \int_0^\infty dL \frac{dn}{dL} \int_0^{z_s} dV \int_{r_-}^{r_+} \frac{rdr}{|M_2(r)|} S(u)C(u) \frac{dN}{dm}(m+2.5\log M(r)) \left[\frac{dN}{dm}(m)\right]^{-1} \quad (5)$$

where $r$ is the coordinate of image 2, $u$ is the source coordinate of the image, $M_2$ is the magnification of image 2, and $M$ is the total magnification of all three images. We selected the gridding in the three dimensions to resolve both the radial and tangential caustics and to keep the fractional errors smaller than 1%. For general lens models with a selection function, no part of this three dimensional integral simplifies, and it must be evaluated separately for each survey quasar. Most approximations found in the literature to simplify the calculation introduce quantitative errors comparable to the differences between $\lambda_0 = 0$ and $\lambda_0 = 1/2$ cosmological models.

### 2.2 Mass Distributions and Dynamical Normalization

We use two mass models for the analysis: the de Vaucouleurs (1948) model, and a softened isothermal sphere model. The de Vaucouleurs model roughly matches the observed surface brightness of galaxies and it is the prototypical model for mass distributions with a constant mass-to-light ratio. The softened isothermal sphere is typical of dark matter dominated models and the inferred mass distribution of spiral galaxies. The hybrid models consisting of a de Vaucouleurs model combined with a softened isothermal sphere halo model used by Maoz & Rix (1993) and Breimer & Sanders (1993) so closely resemble a single softened isothermal sphere model that we do not treat them separately.

The density profile of the softened isothermal sphere is

$$\rho = \frac{\sigma_{DM}^2}{2\pi G(r^2 + s^2)} \quad (6)$$

and it deflects light rays by

$$\frac{\partial \phi}{\partial r} = b \frac{(r^2 + s^2)^{1/2} - s}{r} \quad (7)$$

where $\sigma_{DM}$ is the one-dimensional velocity dispersion of the dark matter, $s$ is the core radius, and $b = 4\pi(\sigma_{DM}/c)^2 D_{LS}/D_{OS}$ (Hinshaw & Krauss 1987). This model was used by Fukugita & Turner (1991), Fukugita et al. (1992), Krauss & White (1992), Maoz & Rix (1993), and Kassiola & Kovner (1993).[1] The strength of the lens is determined by the ratio $\beta = s/b$, and the lens produces multiple images if $\beta > 1/2$. The cross section of the model can be computed analytically (Hinshaw & Krauss 1987), and *under the assumption of a constant comoving core radius* the cross section can be integrated to compute the optical depth (Krauss & White 1992).

---

[1] The other softened isothermal model found in the literature is the two-dimensional lensing potential $\phi = b(r^2 + s^2)^{1/2}$ (Blandford & Kochanek 1987, Kochanek 1991, Wallington & Narayan 1993). Its lensing properties are virtually indistinguishable from the model of equations (6) and (7) if the Blandford and Kochanek (1987) core radius is twice the Hinshaw & Krauss (1987) core radius.

We assume the velocity dispersion $\sigma_{DM}$ is related to the luminosity $L$ by a "Faber-Jackson" law of the form $\sigma_{DM} = \sigma_{DM*}(L/L_*)^{1/\gamma}$, in analogy to the Faber-Jackson (1976) law for the central velocity dispersions of E/S0 galaxies. For a singular isothermal mass model, Kochanek (1994) found that $\sigma_{DM*} = 225 \pm 10$ km s$^{-1}$ and $\gamma = 4.2 \pm 0.2$, compared to $\sigma_{c*} = 228 \pm 14$ km s$^{-1}$ and $\gamma = 3.4 \pm 0.3$ for the central velocity dispersions. We adopt broader uncertainties of $\sigma_{DM*} = 225 \pm 22.5$ km s$^{-1}$ and $\gamma = 4.0 \pm 0.5$ for the remainder of the paper to compensate for systematic errors in the dynamical models. The fact that $\sigma_{DM*}/\sigma_{c*} \simeq 1$ demonstrates that the simple dynamical model with $\sigma_{DM} = (3/2)^{1/2}\sigma_c$ introduced by Turner, Ostriker, & Gott (1984) and used by Turner (1990), Fukugita & Turner (1991), Fukugita et al. (1992), Krauss & White (1992), Wallington & Narayan (1993), Maoz & Rix (1992), and Kassiola & Kovner (1993) is incorrect. Using the Turner, Ostriker, & Gott (1984) dynamical model overestimates the number of lenses by 125% and overestimates their separations by 50%.

It is increasingly clear that E/S0 galaxies are effectively singular. HST observations show cores with surface brightness distributions approaching $R^{-y}$ with $0 \lesssim y \lesssim 1$ (see Tremaine et al. 1994). This range matches that between the Hernquist (1990) and Jaffe (1983) models, where the Jaffe (1983) model is the same as a singular isothermal sphere in its inner regions.[2] The almost uniform absence of central images in the observed lenses (Wallington & Narayan 1993, Kassiola & Kovner 1993) and models of individual lenses (e.g. Kochanek 1995) imply core radii smaller than $100h^{-1}$ pc for $L_*$ galaxies.[3] This suggests that earlier concerns about the effects of finite core sizes on lens statistics (e.g. Blandford & Kochanek 1987, Kochanek & Blandford 1987, Hinshaw & Krauss 1987, Kochanek 1991, Krauss & White 1992) were overstated. Nonetheless, we consider models with core radii, where $s_*$ is the core radius for an $L_*$ galaxy, and $s/s_* = (\sigma/\sigma_*)^x$ with $x = 4.8$. For a "Faber-Jackson" exponent $\gamma = 4$ this corresponds to $s/s_* \propto (L/L_*)^{1.2}$ and matches the observed scaling of the effective radius with luminosity (Fukugita & Turner 1991).

The second model we consider is the de Vaucouleurs (1948) model, as an example of a constant mass-to-light ratio model. The lens deflection produced by the de Vaucouleurs model is

$$\frac{\partial \phi}{\partial r} = \frac{4Gm}{c^2 R_e}\frac{D_{LS}}{D_{OS}}g(R/R_e) \quad \text{where} \quad g(x) = \frac{1}{x}\frac{\int_0^x dx\, x \exp(-kx^{1/4})}{\int_0^\infty dx\, x \exp(-kx^{1/4})} \quad \text{and} \quad k = 7.67. \quad (8)$$

The characteristic deflection scale is

$$b_v = \frac{4Gm}{c^2 R_e}\frac{D_{LS}}{D_{OS}} = 3\rlap{.}''9 \left[\frac{m}{10^{11}h^{-1}M_\odot}\right]\left[\frac{h^{-1}\text{kpc}}{R_e}\right]\frac{D_{LS}}{D_{OS}} \quad (9)$$

for total mass $m$ and effective radius $R_e$. The deflection function peaks at $R_{max} = 0.164 R_e$ with $\max(\phi_{,r}) = 0.74 b_v$. The effective radius of E/S0 galaxies varies with luminosity as $R_e =$ $R_{e*}(L/L_*)^a$ with $R_{e*} = (4 \pm 1)h^{-1}$ kpc and $a = 1.2 \pm 0.2$ where $H_0 = 100h$ km s$^{-1}$ Mpc$^{-1}$ (e.g. Kormendy & Djorgovski 1989, Rix 1991). The average blue mass-to-light ratio varies as $(M/L)_B = (M/L)_{B*}(L/L_*)^b$, where $(M/L)_{B*} = (10 \pm 2)h$ and $b = 0.25 \pm 0.10$ in dynamical models assuming constant mass-to-light ratios (van der Marel 1991, Kormendy & Djorgovski 1989). With $M_*(B_T^0) = -19.9^{-0.2}_{+0.4}$ (Efstathiou, Ellis, & Peterson 1988, Loveday et al. 1992) the deflection scale becomes $b_v = (1\rlap{.}''3 \pm 0\rlap{.}''4)(L/L_*)^{0.05 \pm 0.22} D_{LS}/D_{OS}$. The magnitude of the deflection scale compared to the observed lens separations is the fundamental reason that Maoz & Rix (1993) and Kochanek (1995) found that the de Vaucouleurs models could not fit the observed image separations.

---

[2] The core radius $s$ of the dark matter need not match the core radius of the luminous matter, but there are dynamical problems if the two differ by any large factor.

[3] Steeply dropping radial profiles like the Hubble profile will have larger, finite core radii, but similar high central surface densities, (e.g. Nair, Narasimha, & Rao 1993).

Maoz & Rix (1993) and Breimer & Sanders (1993) also examined hybrid models by adding an isothermal halo with a large core radius to the de Vaucouleurs model. When these models are examined in detail, the hybrid model so closely mimics a softened but nearly singular isothermal sphere that there is little point in making the artificial division into a luminous mass distribution and a dark matter halo with the resulting multiplicity of parameters. In the hybrid models the halo determines the asymptotic deflection and the inner de Vaucouleurs model determines the effective core radius. For example, the deflection produced by a de Vaucouleurs model inside the peak at $R_{max} = 0.164 R_e$ matches that of an isothermal sphere with asymptotic deflection equal to the peak deflection and a core radius of $s = 0.0072 R_e = 30 h^{-1}(L/L_*)^{1.2}$ parsecs. With the addition of a halo the dynamical normalizations must be recomputed (this was done by Breimer & Sanders (1993) and Franx (1993) but not by Maoz & Rix (1993)) with the unsurprising result that the normalizations match those derived for isothermal spheres by Kochanek (1993, 1994).

We want to use lens models that are dynamically consistent, or at least to know when the models capable of fitting the lens data are inconsistent with the requirements of stellar dynamics. Where the dynamical normalization is unknown, we fit constant isotropy dynamical models to the sample of 37 E/S0 galaxies from van der Marel (1991). If $\nu$ is the luminosity density of the galaxies found by deprojecting the observed surface density, $\beta = 1 - \sigma_\theta^2/\sigma_r^2$ is the constant isotropy, and $M(r)$ is the mass, then the radial velocity dispersion is

$$\nu \sigma_r^2 = r^{-2\beta} \int_r^\infty \frac{GM\nu}{r^2} r^{2\beta} dr \qquad (10)$$

and the line of sight velocity dispersion is

$$\Sigma v_{los}^2 = 2 \int_0^\infty dz \, \nu \sigma_r^2 \left(1 - \beta \frac{R^2}{r^2}\right) \qquad (11)$$

where $z$ is the coordinate along the line of sight, $R$ is the projected radius, and $\Sigma$ is the observed surface brightness of the galaxy (Binney & Tremaine 1987).[4] For the purposes of this paper we fit isotropic models ($\beta = 0$) but broaden the errors on the dynamical variables by approximately a factor of two.

---

[4] Note the typographical errors in equations (2) and (3) of Kochanek (1994). Equation (2) should read $\nu \sigma_r^2$ and equation (3) should read $\int_{-\infty}^\infty$.

## 2.3 Statistical Analysis

Given a lens model, the number counts of galaxies, the properties of the galaxies, and the lens survey selection function, we compute the probability $p_i$ that quasar $i$ is lensed or the probability $p_i(\Delta\theta_i)$ that quasar $i$ is lensed and has image separation $\Delta\theta_i$ using equation (5). If the sample contains $N_U$ unlensed quasars, $N_L$ lensed optical quasars, and $N_R$ radio lenses, the likelihood function for the data $d$ given a fixed set of model parameters $\xi$ is

$$\ln L(d|\xi) = -\sum_{i=1}^{N_U} p_i + \sum_{j=1}^{N_L} \ln p_j(\Delta\theta_j) + \sum_{k=1}^{N_R} \ln\left(\frac{p_k(\Delta\theta_k)}{p_k}\right). \tag{12}$$

We expanded the logarithm of the probability that a quasar is unlensed $\ln(1-p_i) \simeq -p_i$ for $p_i \ll 1$, and we include only the relative probability that a radio lens has separation $\Delta\theta_k$ rather than the absolute probability that it was lensed by using the ratio $p_k(\Delta\theta_k)/p_k$. The various probabilities include all the model parameters and survey selection function models. We can add other aspects of the configuration of the lenses to the likelihood such as the lens galaxy redshift, the detectability of odd/central images, and the image morphology if desired.

The model parameters also have uncertainties that are not included in any of the existing cosmological limits. For most parameters we have estimates of their mean values and dispersions, so we can define the prior probability $P(\xi_j)$ for parameter $\xi_j$. The prior probabilities are assumed to be normal distributions for signed variables and log-normal distributions for positive definite variables. Using Bayes theorem, the probability of the parameters given the data is then

$$P(\xi|d) = \frac{\Pi_j P(\xi_j) L(d|\xi)}{\int \Pi_j d\xi_j \Pi P(\xi_j) L(d|\xi)}. \tag{13}$$

The denominator of the expression merely normalizes the total probability to be unity. We use the Bayesian analysis because it provides a clear, formal approach for including the parameter uncertainties, for defining error bars on non-Gaussian distributions, and for projecting out nuisance variables. The probability distribution for any subset of the parameters is found by integrating the probability distribution over the unwanted variables.

We usually reduce the distribution to a two dimensional parameter space and present likelihood ratio contours in this subspace. Usually we examine the space of the cosmological model and the dynamical variable. The 68%, 90%, 95%, and 99% confidence limits on one parameter are the extrema of the likelihood ratio contours at 61%, 26%, 14% and 3.6% of the peak likelihood. The Bayesian limit on one variable is found by marginalizing the distribution to a single variable and finding the limits enclosing the relevant fraction of the probability.

## 2.4 Number Counts of Galaxies

We divide the number counts of galaxies into an E/S0 class and spirals, with comoving densities fitted by a Schechter (1976) function,

$$\frac{dn}{dL} = \frac{n_*}{L_*} \left(\frac{L}{L_*}\right)^\alpha \exp(-L/L_*).$$

We use an average comoving density of galaxies of $n_* = (1.40 \pm 0.17) h^3$ Mpc$^{-3}$ (Loveday et al. 1992) where $H_0 = 100 h$ km s$^{-1}$ Mpc$^{-1}$. The division of the luminosity function by galaxy type is taken from Marzke et al. (1994), renormalized to the total galaxy density found by Loveday et al. (1992). The Loveday et al. (1992) sample is too deep to type the fainter ellipticals (Marzke et al. (1994)), while the total density in the Marzke et al. (1994) sample is too high because of local structures. With these assumptions the comoving densities of E/S0 and spiral galaxies are

$$n_e = (0.61 \pm 0.21) h^3 \times 10^{-2} \text{ Mpc}^{-3} \quad \text{and} \quad n_s = (0.79 \pm 0.26) h^3 \times 10^{-2} \text{ Mpc}^{-3}. \quad (14)$$

We adopt a common slope $\alpha = -1.00 \pm 0.15$ for both galaxy types (Marzke et al. 1994), which also agrees with the total slope found by Loveday et al. (1992). These estimates supersede the (statistically and systematically) less accurate model of Fukugita & Turner (1991) based on the earlier number counts of Efstathiou, Ellis & Peterson (1988) and the morphological type ratios of Postman & Geller (1984). While the error bars in the newer surveys are more reliable, they are also broader. The quoted error estimates are all one standard deviation errors. The shortcoming of this error model is that the values of $\alpha$ and $n_*$ must be correlated, but neither Marzke et al. (1994) nor Loveday et al. (1992) discuss the parameter uncertainties in the full three-dimensional space of $\alpha$, $n_*$ and $L_*$. The absolute blue magnitude of an $L_*$ E/S0 galaxy is $M_*(B_T^0) = -19.9^{-0.2}_{+0.4}$ magnitudes, which corresponds to luminosity $L_* = 1.4^{+0.3}_{-0.5} \times 10^{10} L_{B\odot}$ (Efstathiou, Ellis, & Peterson 1988, Loveday et al. 1994). When we calculate the selection effects from the galaxy luminosity (§7.1) and the probability of measuring a lens galaxy redshift (§8), we use the $k$-corrections model of Coleman, Wu, & Weedman (1980) to compute the apparent magnitude of the lens galaxy. The model does not include evolutionary corrections, and it is probably accurate only to within one magnitude.

## 2.5 Number Counts of Quasars

The number counts model for the quasars is critical to the statistical analysis because magnification bias (Gott & Gunn 1974, Turner 1980) has an enormous effect on lens probabilities in bright quasar surveys. No realistic calculation of lens probabilities can neglect its effects. In this section we compare the various number counts models used in lensing calculations.

Turner (1990), Fukugita & Turner (1991), and Kochanek (1991, 1993) used a simple two power law model with a fixed apparent break magnitude developed by Turner (1990) based on the compilation of data by Hartwick & Schade (1990). This model has a break at $m_0 = 19.15$ B mags, a bright end slope $\alpha = 0.86$ and a faint end slope $\beta = 0.28$ where

$\log_{10} dN/dm = \alpha(m - m_0) + c$ for $m < m_0$ and $\log_{10} dN/dm = \beta(m - m_0) + c$ for $m > m_0$. We are uninterested in the value of the constant $c$ for lens calculations. The bright slope $\alpha$ in the Turner (1990) model is the slope for the apparent magnitude number counts of all bright, UV-excess selected quasars with $z \lesssim 2.2$ (Hartwick & Schade 1990). By averaging over the redshifts of the quasars, this model significantly underestimates the true slope of the luminosity function.

Wallington & Narayan (1992) and Kassiola & Kovner (1993) use a quasar luminosity function modeled on that of Boyle et al. (1988). In this model the luminosity function is

$$\frac{dN}{dm} \propto \left[ 10^{-\alpha(m-m_0)} + 10^{-\beta(m-m_0)} \right]^{-1} \tag{15}$$

where the bright end slope is $\alpha = 1.12 \pm 0.06$, and the faint end slope is $\beta = 0.18 \pm 0.08$. The break magnitude $m_0$ evolves with redshift based on the Boyle et al. (1988) model for redshifts less than three, and with a correction of $-0.54(z-3)$ magnitudes above a redshift of three to fit the higher redshift surveys (Wallington & Narayan 1993).[5] The Snapshot Survey and Maoz & Rix (1993) used a two power law model in absolute magnitude based on the luminosity function of Boyle et al. (1987). The bright end slope is $\alpha = 1.04$, the faint end slope is $\beta = 0.08$, and the evolutionary model for $m_0$ is identical to that of Wallington & Narayan (1993).

The complexity of the absolute magnitude models hides the fundamental simplicity of the evolution in apparent magnitude. Between the redshifts of 1 and 3, the apparent magnitude of the break is $\simeq 19$ B magnitudes. Below a redshift of one and above three we allow the break magnitude to evolve as

$$m_0(z) = \begin{cases} m_0 + (z - 1.0) & z < 1 \\ m_0 & 1 < z < 3 \\ m_0 - 0.7(z-3) & 3 < z \end{cases} \tag{16}$$

This evolution model matches the Wallington & Narayan (1993), Boyle et al. (1988, 1990), and Maoz & Rix (1993) models to higher accuracy than required by the statistical uncertainties. We fit this evolution model and the broken power law functional form to the quasar luminosity function data in Hartwick & Schade (1990) for $z > 1$ to estimate the three parameters $\alpha$, $\beta$, and $m_0$.

Figure 1 shows the projected likelihood functions for the three parameters, with crosses at the Fukugita & Turner (1991), Maoz & Rix (1993), and Wallington & Narayan (1993) models. The best models with the simple apparent magnitude evolution model in equation (16) are slightly better than the more complicated evolution models used by Wallington & Narayan (1993) and Maoz & Rix (1993), although the improvements are not significant.

---

[5]It appears that a conversion between $H_0 = 50$ km s$^{-1}$ Mpc$^{-1}$ used by Boyle et al. (1988) and $H_0 = 100$ km s$^{-1}$ Mpc$^{-1}$ used by Wallington & Narayan (1993) was lost in their conversion to apparent magnitudes. This implies that figures as a function of apparent magnitude in Wallington & Narayan (1993) and Kassiola & Kovner (1993) (who used the Wallington & Narayan (1993) model) must be shifted by 1.5 magnitudes.

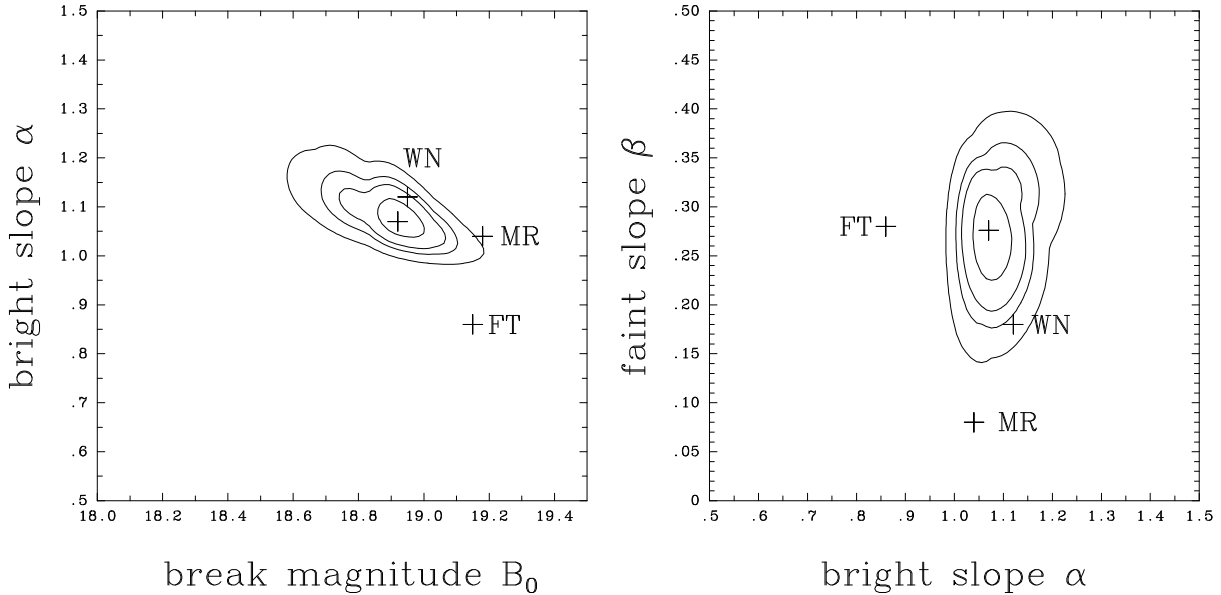

FIG. 1.–Likelihood contours for the quasar number counts model. In each two dimensional subspace the third variable was optimized. The three parameters are the bright slope $\alpha$, the faint slope $\beta$, and the break magnitude $m_0$. Contours are shown at the 68%, 90%, 95%, and 99% confidence limits for one variable in the likelihood ratio. The crosses mark the best fit, the Fukugita & Turner (1991) model (FT), the Wallington & Narayan (1993)/Boyle et al. (1988) model (WN), and the Maoz & Rix (1993)/Boyle et al. (1987) model (MR). The FT model has too flat a bright slope to be consistent with the data, the MR model has too flat a faint slope to be consistent with the data, and the WN model is consistent with the data.

The best model has $\alpha = 1.07 \pm 0.07$, $\beta = 0.27 \pm 0.07$, and $m_0 = 18.92 \pm 0.16$ B mags (90% confidence error bars) with $\chi^2 = 63.6$ for $N_{dof} = 62$ degrees of freedom. A one standard deviation change in the $\chi^2$ from $\chi^2 = N_{dof}$ is $\Delta\chi^2 = 11$ for 3 parameters. The bright end slope of $\alpha = 0.86$ in the Fukugita & Turner (1991) model is in complete disagreement with the data – the likelihood ratio is $\log(L/L_{max}) = -12.1$. The best model with the Wallington & Narayan (1993) slopes has $m_0 = 18.95$ and $\log(L/L_{max}) = -0.95$. This is within the 90% confidence interval, and our best fit is no better than the full Wallington & Narayan (1993) model using the Boyle (1988) functional form rather than the broken power law. The best model with the Maoz & Rix (1993)/Boyle et al. (1987) slopes has $m_0 = 19.18$ and $\log(L/L_{max}) = -2.5$. It can be rejected with 99% confidence because the faint slope of $\beta = 0.08$ is inconsistent with the data and with later analyses (Boyle et al. 1988, 1990). We will use our best fit model ($\alpha = 1.07$, $\beta = 0.27$, and $m_0 = 18.92$ B mags) and the evolution model for $m_0$ in equation (16). We let the quasar density be constant beyond $m_c = 30$ B magnitudes, and we use an average $B - V$ color of 0.2 mag as suggested by Maoz et al. (1993ab).

| Table 1: Summary of Lens Data | | | | | | | |
|---|---|---|---|---|---|---|---|
| Lens | $z_s$ | $z_l$ | $m_s$ mag | $m_l$ mag | $m_{i0}$ mag | $\Delta\theta$ arcsec | Comments |
| 1208+1011 | 3.80 | | 17.9 | | 20.4 | 0.45 | |
| H1413+117 | 2.55 | | 17.0 | | 19.5 | 1.2 | |
| LBQS 1009−0252 | 2.75 | | 17.9 | | 20.0 | 1.6 | |
| PG 1115+080 | 1.72 | 0.29 | 16.1 | 19.8 | 21.2 | 2.2 | |
| 0142−100 | 2.72 | 0.49 | 16.8 | 19.0 | 21.0 | 2.2 | |
| 0957+561 | 1.41 | 0.36 | 16.7 | 18.5 | − | 6.1 | Not Used |
| B 0218+356 | | 0.68 | | 20 | − | 0.35 | |
| MG 0751+2716 | | | | 19 | − | 0.9 | |
| B 1422+231 | 3.62 | | 16.5 | | 19.0 | 1.3 | |
| CLASS 1600+434 | 1.61 | | 20.0 | | 19.0 | 1.4 | |
| MG 1549+3047 | | 0.11 | 23 | 16 | − | 1.7 | |
| B 1938+666 | | | (23) | | − | 1.8 | |
| MG 1654+1346 | 1.74 | 0.25 | | 19.0 | 21.0 | 2.1 | |
| MG 0414+0534 | 2.64 | | 21.8 | I'=21.4 | − | 2.1 | |
| MG 1131+045 | | | | (22) | − | 2.1 | |
| CLASS 1608+656 | 2.30 | (0.63) | (20) | (20) | 21.0 | 2.1 | |
| 2016+112 | 3.27 | (1.01),? | i=21.8 | i=21.9 | − | 3.8 | Not Used |

NOTES: The source and lens redshifts are $z_s$ and $z_l$, the source and lens magnitudes are $m_s$ and $m_l$, and the image separation or the diameter of the tangential critical line is $\Delta\theta$. The magnitude $m_{i0}$ is our estimate of how bright the lens galaxy has to be for the lens redshift to be measurable for the systems used in §8. The first six lenses are in the optical surveys, and the last eleven lenses are in the radio surveys. An entry in parenthesis means low accuracy or substantial uncertainty. The magnitudes are R magnitudes unless otherwise specified. See Surdej & Soucail (1994) for more details and the original references.

### 2.6 Observational Data & Selection Functions

We use the optical lens surveys by Crampton et al. (1992), Yee et al. (1992), the Snapshot Survey (Bahcall et al. 1992, Maoz et al. 1992, 1993ab), the ESO/Liège Survey (Surdej et al. 1993), the NOT survey (Jaunsen et al. 1995), the HST GTO survey (see Falco 1994), and the FKS Survey (Kochanek et al. 1995) but include only the quasars with $z > 1$. We use the conservative evaluation of the ESO/Liège Survey candidates by Kochanek (1993), and the preliminary FKS selection function. The combined sample contains 862 quasars and six lenses. Table 1 summarizes the observational data. We do not use 0957+561 in the analysis because the lens consists of both a cluster and a galaxy. Kochanek (1995) and Wambsganss et al. (1995) examine the statistics of wide separation lenses like 0957+561.

In addition to the optical samples we include the image separations of the radio lenses found in the MIT/Greenbank (MG) Survey (see Burke et al 1992, Patnaik 1994), the Jodrell Bank Survey (see Patnaik et al. 1992, Patnaik 1994), and the CLASS survey (Jackson et al. 1995, Myers et al. 1995). Because the redshift and luminosity functions of these radio surveys are incompletely known, it is impossible to calculate the absolute probability of gravitational lensing in these samples reliably. We can, however, estimate the relative likelihoods of finding the observed image separations with much weaker uncertainties from the effects of the magnification bias on the likelihood of finding a particular separation. There are eleven of these radio lenses and Table 1 summarizes the observational data. We used all these lenses except 2016+112 in the analysis. The lens 2016+112 is probably a complicated hybrid lens involving two galaxies at different redshifts. Where the source redshift is unknown, it is set to $z_s = 2$. The optical magnitudes are set to $m = 19$ for the magnification bias calculation. The bias near $m = 19$ is comparable to that expected for the radio samples. As discussed by Kochanek (1993) the source redshift has little effect on the image separations, particularly in flat cosmological models.

### 3  Isothermal Lenses and Parameter Uncertainties

Independent of any systematic problems in the statistical model, we must consider how the uncertainties in the model parameters affect the cosmological conclusions. The known statistical uncertainties in galaxy number counts, quasar number counts, and dynamical relations have not been included in any of the full analyses of lens statistics. Kochanek (1993) included the uncertainties in the dynamical normalization for the lens galaxies and examined the effects of the other variables in a particular cosmological model, while Maoz & Rix (1993) examined "conspiracy" models in which parameters were biased to increase the likelihood of a large cosmological constant. In this section we derive the cosmological limits including the uncertainties of the input parameters. We use the only known lens model consistent with lens statistics, galaxy dynamics, and individual lens models, the nearly singular isothermal sphere, with a core radius of $s_* = 10h^{-1}$ pc. This model also mimics the Breimer & Sanders (1993), Franx (1993), and Maoz & Rix (1993) models of an E/S0 galaxy embedded in a dark matter halo. All later calculations will automatically include these uncertainties.

#### 3.1  Galaxy Number Counts and Dynamical Relations

The variables relating galaxy number counts and the isothermal lens model are the number density of E/S0 galaxies $n_e = (6.1 \pm 2.1)h^3 \times 10^{-3}$ Mpc$^{-3}$, the Schechter (1976) function slope $\alpha = -1.0 \pm 0.15$, the Faber-Jackson (1976) exponent $\gamma = 4.0 \pm 0.5$, and the velocity dispersion of the dark matter for an $L_*$ galaxy, $\sigma_{DM*} = 225 \pm 22.5$ km s$^{-1}$ (Kochanek 1994). The prior probability distributions for the variables are assumed to be the log-normal distribution for the galaxy number density and the normal distribution for the other variables with the stated one-standard deviation uncertainties. If $\langle D_{OS}^3 \rangle$ is the sample average of the cube of the proper motion distance to the quasars, the expected number of lenses varies as $n_e \langle D_{OS}^3 \rangle \sigma_{DM*}^4 \Gamma[1 + \alpha + 4/\gamma]$ and the mean image separation varies as $\sigma_{DM*}^2 \Gamma[1 + \alpha + 6/\gamma]/\Gamma[1 + \alpha + 4/\gamma]$ (see Kochanek 1993). Figure 2 shows the two-dimensional likelihood contours for each of the four variables and the matter density $\Omega_0$ in a flat universe after marginalizing the original five dimensional likelihood function over the other three variables.

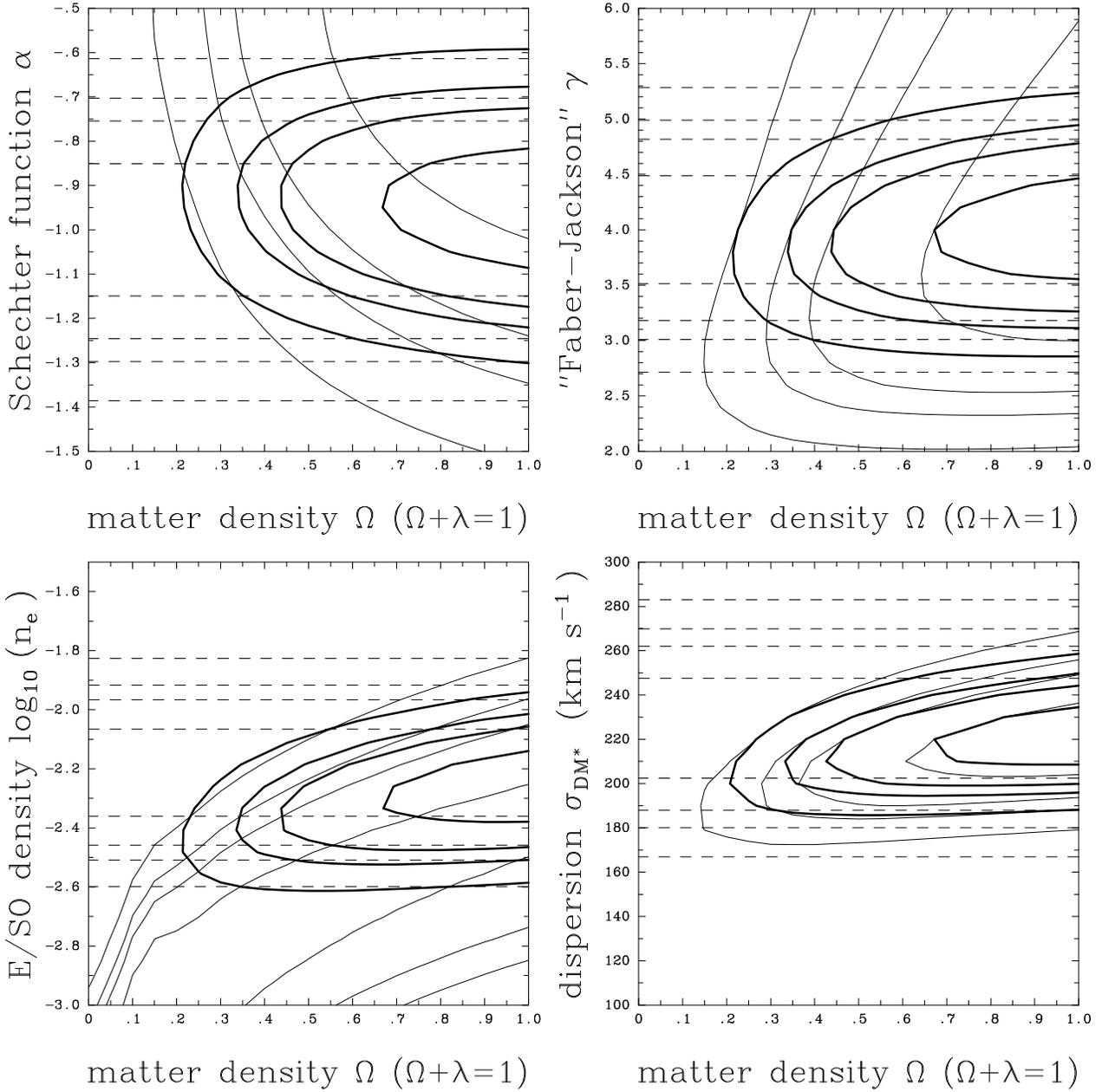

FIG. 2.—Likelihood contours showing the dependence of the cosmological limits on the parameters determining the galaxy number counts and dynamical relations. Contours are shown at the 68%, 90%, 95%, and 99% confidence limits for one variable in the likelihood ratio. The dashed lines show the prior probability distribution for the variable ($\alpha$, $\gamma$, $n_e$, or $\sigma_{DM*}$), the light solid lines show the limits using only the lensing data (although including the prior probability distributions for the other three variables), and the heavy solid lines show the joint limits including the lensing data and the prior probability distribution. The galaxy density is in units of $h^3$ Mpc$^{-3}$.

With the inclusion of all the statistical uncertainties (artificially broadened uncertainties for $\gamma$ and $\sigma_{DM*}$) the two standard deviation upper limit on the cosmological constant is $\lambda_0 < 0.66$. The best fit models with $\Omega_0 \simeq 1$ are consistent with the external estimates of the input parameters. In particular, the dynamical estimates of $\sigma_{DM*}$ from Kochanek (1994) exactly match the best estimates from fitting the lens separations unless $\lambda_0 \sim 1$. The existence of a $(3/2)^{1/2}$ correction for the dark matter velocity dispersion is supported neither by galaxy dynamics nor the observed separations of gravitational lenses. The best fit estimate is $\sigma_{DM*} = 220 \pm 20$ km s$^{-1}$, and it is well constrained because the average image separation is a strong function of the velocity dispersion, $\Delta\theta \propto \sigma_{DM*}^2$.

The lensing data is least able to differentiate between changes in the galaxy number density $n_e$ and the cosmological model. Because the image separation distribution for the singular isothermal sphere does not depend on cosmology, only the variation of the lens probability with redshift differentiates between the galaxy number density and cosmology. With only five optical lenses (the radio lenses do not matter here) there is too little data to measure the variation with redshift. This is a fundamental uncertainty, and any other systematic problem that can mimic changes in $n_e$ is similarly difficult to constrain. For example, using dust to make some fraction of the E/S0 galaxies completely opaque at all redshifts is the same as changing $n_e$.

The Schechter (1976) function slope $\alpha$ controls the relative numbers of low and high mass galaxies. The lens statistics are dominated by the number of $\gtrsim L_*$ galaxies as long as $\alpha \gtrsim -1$, and in this regime the results will be insensitive to the exact value of the exponent. When $\alpha \lesssim -1$ the number of lenses produced by the small galaxies rises rapidly, so there are strong restrictions on $\alpha \lesssim -1$. In reality there is a correlation between the value of $n_e$ and the value of $\alpha$. Unfortunately the studies of galaxy number densities (Marzke et al. 1994, Loveday et al. 1992, Efstathiou et al. 1988) give only the value of $n_e$ and its uncertainties for the best fitting value of $\alpha$, so we are unable to include it in the error model.

The "Faber-Jackson" exponent is weakly constrained by the lensing data to lie near the standard values of $\gamma \simeq 4$ with strong constraints on $\gamma < 4$. The expected number of lenses is minimized for $\gamma \simeq 2.8$ (for $\alpha = -1$), and there the likelihood contours reach their peak values of the cosmological constant. The exponent $\gamma$ and $\sigma_{DM*}$ are correlated, but we chose to use broad uncorrelated error bars for the two variables separately rather than a detailed correlation matrix that was dominated by systematic uncertainties in the dynamical modeling.

### 3.2 Quasar Number Counts

No study of the lens surveys has examined the sensitivity of the cosmological results to the quasar number counts model. Kochanek (1993) examined how well the lensing model could estimate the values of the individual parameters of the number counts model for a fixed cosmological model and found that the lensing data was better fit by a steeper slope for the number counts of bright quasars than in the Fukugita & Turner (1991) model. Since there is no convenient way of storing intermediate results to allow a multi-dimensional parameter study as used for the distribution of lens galaxies, we restricted ourselves to a small Monte Carlo study. We selected 12 sets of number counts parameters from our fits to the Hartwick & Schade (1990) compilation, as well as the Fukugita & Turner (1991) number counts model and the Maoz & Rix (1993)/Boyle et al. (1987) number counts model, for a total of 14 different models. The probability of fitting the separations of the radio lenses was kept fixed to avoid introducing any biases in the models.

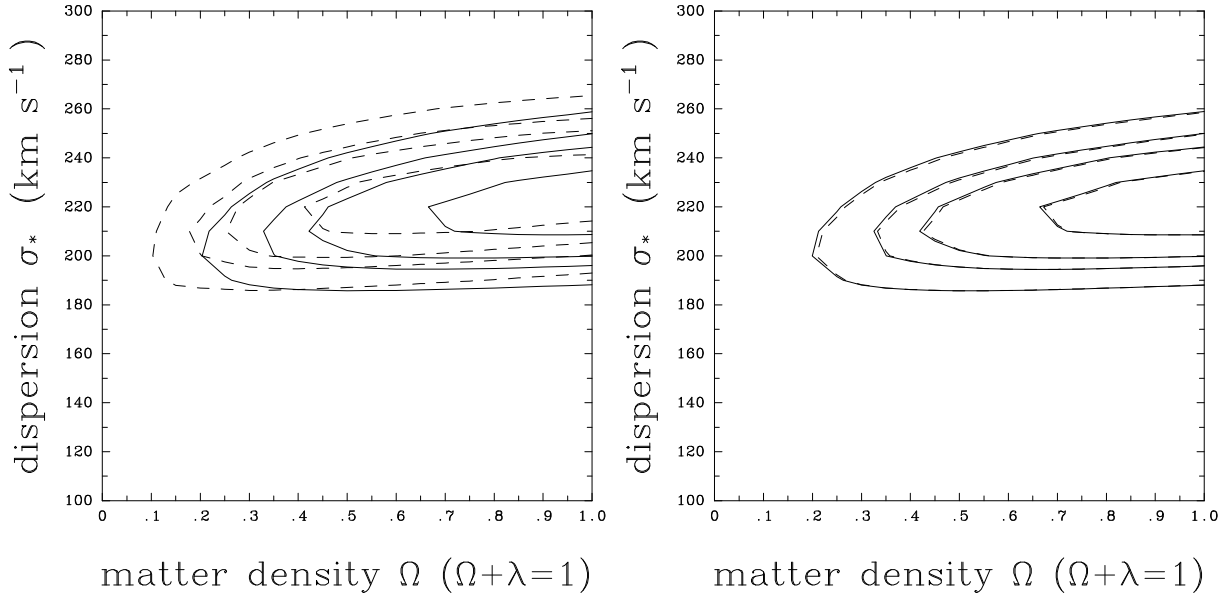

FIG. 3.—Likelihood contours showing the dependence of the cosmological limits on the quasar number counts model. Contours are shown at the 68%, 90%, 95%, and 99% confidence limits for one variable in the likelihood ratio. In the left panel the dashed contours are for the Fukugita & Turner (1991) number counts model and the solid contours are for the Maoz & Rix (1993)/Boyle et al. (1987) number counts model. In the right panel, the dashed contours are for our best fit number counts model, and the solid contours are an average of the best fit model plus eleven other models randomly selected from the probability distribution of the fits to the quasar number counts. The likelihood distributions include the uncertainties in $\sigma_{DM*}$, $n_e$, $\alpha$, and $\gamma$.

Figure 3 shows the likelihood contours in the space of $\sigma_{DM*}$ and $\Omega_0$ including the prior probability distributions for $n_e$, $\alpha$, $\gamma$, and $\sigma_{DM*}$. The best fit model derived in §2.5, the Maoz & Rix (1993) model, and the Wallington & Narayan (1993) model all produce similar results, and averaging over the small Monte Carlo sampling of the distribution has little effect on the cosmological conclusions. The Fukugita & Turner (1991) model, however, produces different cosmological limits because of its flatter bright quasar slope. The Fukugita & Turner (1991) number counts model give 68%, 90%, 95% and 99% confidence limits on the cosmological constant of $\lambda_0 \lesssim 0.58$, 0.75, 0.84, and 0.90 respectively while the other three models give $\lambda_0 \lesssim 0.33$, 0.57, 0.66, and 0.78 respectively. The underestimate of the magnification bias by the Fukugita & Turner (1991) number counts model is the reason that Kochanek (1993) found weaker limits on the cosmological constant than Maoz & Rix (1993).

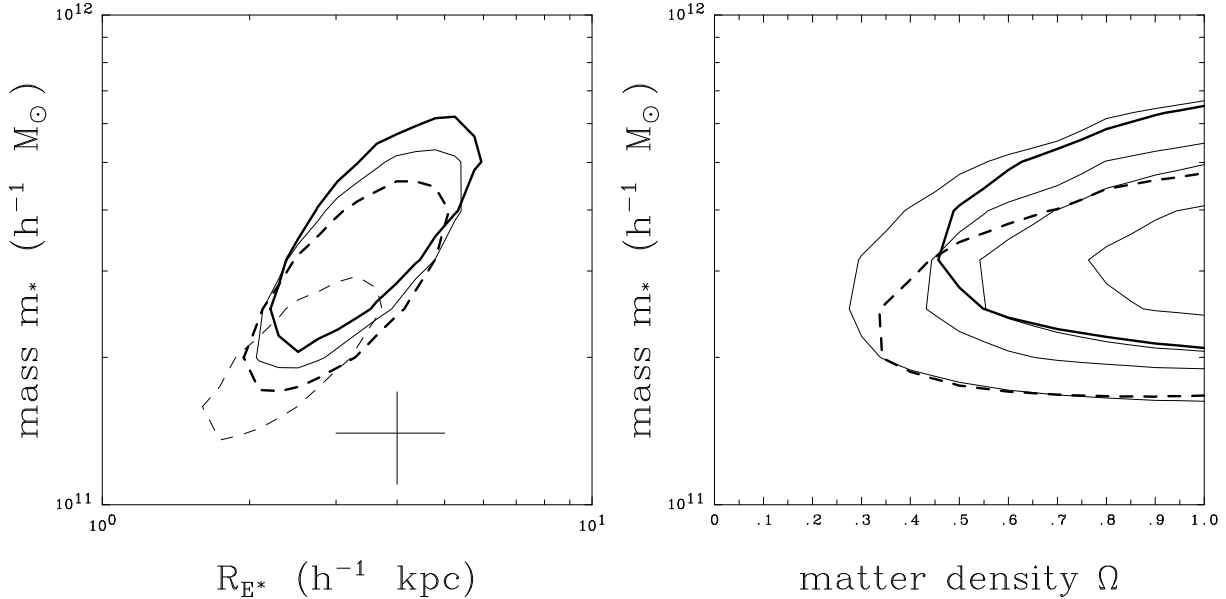

Fig. 4.–The left panel shows the 95% confidence contour for the the mass $m_*$ and effective radius $R_{e*}$ of an $L_*$ galaxy for $\Omega_0 = 1$ (light solid) and $\lambda_0 = 1$ (light dashed) when $a' = 1$. The dashed (solid) heavy solid lines show the 95% confidence contours for $\Omega_0 = 1$ and $a' = 0.8$ ($a' = 1.2$). The likelihoods include the observed log-normal prior probability distribution for $R_{e*}$ and they are drawn relative to the peak for each model. The cross shows the observational uncertainties for $R_{e*}$ and $m_*$ in constant mass-to-light ratio models. The right panel shows the likelihood contours for the mass of the galaxy and the cosmological model. The uncertainties in the galaxy number counts ($\alpha$ and $n_e$), the normalization of the effective radius $R_{e*}$, the mass scale $m_*$ and the scaling exponent of the mass-to-light ratio $b$ are included in the error bars. The light solid lines show the 68%, 90%, 95%, and 99% confidence intervals for one parameter in the likelihood ratio when $a' = 1$, and the dashed (solid) heavy lines show the 95% confidence interval for $a' = 0.8$ ($a' = 1.2$).

Because the spread in the maximum likelihoods is only a factor of two, we cannot choose between the luminosity function models. The sample has one bright (PG 1115+080), one faint (1208+101), and three intermediate (0142−100, LBQS 1009−0252, H 1413+117) magnitude quasars. The likelihoods are similar because the gains (or losses) in the probability of PG 1115+080 are compensated by a losses (or gains) in the probability of 1208+101. Thus the probability for the parameters of the quasar luminosity function are determined by the prior probability distribution for fitting the observed quasar number counts. We use only the best fit model from §2.5 for the remainder of the paper.

# 4 DE VAUCOULEURS MODELS

From both lens models (Kochanek 1995) and lens statistics (Maoz & Rix 1993, Kochanek 1995) we know that the de Vaucouleurs mass model for lens galaxies with standard mass-to-light ratios is unable to fit lens data. One solution is to add a halo to the de Vaucouleurs model (Maoz & Rix 1993, see also Breimer & Sanders 1993), but these galaxy + halo models so closely resemble softened isothermal sphere models that we will not treat them separately. In this section we explore the properties of the de Vaucouleurs models to estimate the galaxy masses required to fit the lens data, and to see if the radically different shape of the de Vaucouleurs model has any effect on the cosmological limits.

The model depends on seven parameters: $\alpha$, $n_e$, $R_e$, $a$, $(M/L)_*$, $b$, and $M_*(B_T)$ (see §2.2). We combine the mass-to-light ratio $(M/L)_*$ with the luminosity scale $M_*(B_T)$ and fit the mass of an $L_*$ galaxy $m_*$ as the fundamental parameter. We can do this because we do not have the covariance matrix for $\alpha$, $n_e$ and $L_*$ in the galaxy number counts models. The mass scales with the luminosity as $m/m_* = (L/L_*)^{1+b}$ where $b$ is the exponent for the scaling of the mass-to-light ratio $(M/L) = (M/L)_*(L/L_*)^b$. We tabulated the lensing probabilities as a function of mass and effective radius, which makes it easy to examine scaling laws of the effective radius with mass ($R_e \propto m^{a'}$ with $a' = a/(1+b)$) but hard to examine ones with luminosity ($R_e \propto L^a$). For our standard models we leave the mass $m_*$ as a free parameter to be determined from the lens data, and average over the uncertainties in the galaxy number counts model ($\alpha$ and $n_e$), the effective radius of an $L_*$ galaxy ($R_{e*} = (4\pm1)h^{-1}$ kpc), and the exponent of the mass-to-light ratio ($b = 0.25 \pm 0.10$) for a fixed value of $a' = a/(1+b) = 0.96 \pm 0.25$. We examine $a' = 0.8$, 1.0, and 1.2, which spans the range for $a$. The average over the value of $R_{e*}$ can be viewed either as including the uncertainties in $R_{e*}$ or as an average over the fundamental plane.

The curious property of the standard scalings for the de Vaucouleurs model is that the peak deflection is proportional to $L^{1+b-a} = L^{0.05\pm0.3}$ compared to $L^{0.5}$ in the isothermal models. The peak deflection is nearly the same for all luminosities, and either an $L_*$ galaxy can generate the largest observed separations or no galaxy can. The peak deflection is proportional to $m/R_e$ and by the virial theorem we expect that $m/R_e \propto \sigma^2$. Observed velocity dispersions have $\sigma^2 \propto L^{0.5}$ so the standard de Vaucouleurs scalings do not match the Faber-Jackson (1976) law because either the slope of the mass-to-light ratio is too shallow, the slope of the effective radius scaling is too steep, or the measured velocity dispersion differs systematically from the virial dispersion.

If we completely ignore the observed effective radii of galaxies and determine what de Vaucouleurs profile would (in abstract) best fit the data, it is a compact ($R_e \simeq 400h^{-1}$ pc), low mass ($m_* \simeq 8 \times 10^{10} h^{-1} M_\odot$) galaxy with $\lambda_0 \lesssim 0.75$ at 95% confidence. When we add the prior probability distribution for the observed effective radii, the best fit models for $\Omega_0 = 1$ are $R_{e*} = 3.0h^{-1}$ kpc and $m_* = 2.5 \times 10^{11} h^{-1} M_\odot$ for $a' = 0.8$, $R_{e*} = 3.3h^{-1}$ kpc and $m_* = 3.2 \times 10^{11} h^{-1} M_\odot$ for $a' = 1.0$, and $R_{e*} = 4.0h^{-1}$ kpc and $m_* = 4.0 \times 10^{11} h^{-1} M_\odot$ for $a' = 1.2$. As $a'$ is lowered, the effective radius and mass decrease. The left panel of Figure 4 shows the 95% confidence regions for the parameters. The models with $\lambda_0 = 1$ do not fit the data nearly as well because the models are driven to low masses and small effective radii to maintain constant image separations while minimizing the number of lenses. The right panel of Figure 4 shows the cosmological limits for $a' = 0.8$, 1.0, and 1.2. In each case the best fit model is $\Omega_0 = 1$, and the 95% confidence upper limits on $\lambda_0$ are $\lambda_0 \lesssim 0.68$, $\lambda_0 \lesssim 0.62$, and $\lambda_0 \lesssim 0.55$. The results for a weighted average of the results for different $a'$ are similar to those for $a' = 1.0$.

In short, we confirm the results of Maoz & Rix (1993) using a larger data set and a more thorough statistical analysis. Constant mass-to-light ratio models of E/S0 galaxies are inconsistent with gravitational lenses unless you reject the dynamically determined estimates of the masses. Including the cosmological uncertainties, the lens data can be fit if $m_* \simeq 3.1^{+2.2}_{-1.3} \times 10^{11} h^{-1} M_\odot$ at 95% confidence. This corresponds to a mass-to-light ratio of $(M/L)_{B*} \simeq 22^{+16}_{-9} h$ at 95% confidence compared to $(M/L)_{B*} = (10 \pm 2)h$ in dynamical estimates (e.g. van der Marel 1991). The cosmological limits when the masses are high enough to fit the observed image separations are nearly identical to the limits we derived in the last section using the isothermal sphere model.

## 5 THE EFFECTS OF A CORE RADIUS

In this section we consider softened isothermal density profiles in more detail to understand the effects of a core radius. The three issues we must treat are the dynamical normalization of models with a finite core radius, the lensing effects of softened isothermal spheres in the presence of magnification bias, and the cosmological effects of introducing a finite core radius.

### 5.1 The Dynamical Normalization of Softened Isothermal Spheres

We need to examine the dynamical normalization because the values of $\sigma_{DM*}$ and $s_*$ are correlated. The core radius reduces the mass near the center of the galaxy and the velocity dispersion must increase compared to the value in a singular model to maintain either the same observed velocity dispersions or the same average image separations. As a model calculation we computed the average line-of-sight velocity dispersion inside one effective radius $R_e$ assuming a Hernquist (1990) model ($\nu \propto r^{-1}(r+a)^{-3}$) for the distribution of the stars with $a \simeq 0.45 R_e$. With the assumption that the velocity dispersion tensor is isotropic ($\beta = 0$ in equation (10)), the dark matter dispersion increases as $\sigma_{DM} \propto 1 + 2(s/R_e)$ with the addition of a core radius. The numerical coefficient in the scaling law is a function of the averaging area, and it rises from 2 to 2.5 if we fix the dispersion inside $R_e/2$ instead of $R_e$.

For a more realistic model we use the 37 galaxies in the van der Marel (1991) sample and fit isotropic dynamical models to each galaxy using equations (10) and (11), assuming the core radius is a constant fraction of the estimated effective radius for each galaxy, and that the velocity dispersion scales as $L/L_* = (\sigma_{DM}/\sigma_{DM*})^4$. The $\chi^2$ surface of the fit to the observed velocity dispersion profiles is shown in Figure 5, and the dashed line is the scaling law estimated from the Hernquist (1990) model. Models with large core radii cannot fit the data because of the contradiction between a homogeneous core and a steeply rising luminosity profile. The formal 95% confidence upper limit on the core radius is $s_* \lesssim 0.08 R_e$ or $s_* \lesssim 300 h^{-1}$ pc for $R_{e*} = 4 h^{-1}$ kpc. While we have greatly oversimplified the dynamical problem and ignored the effects of varying the isotropy of the velocity ellipsoid, the basic physical effect that the velocity dispersion must rise as the core radius increases is independent of any complications.

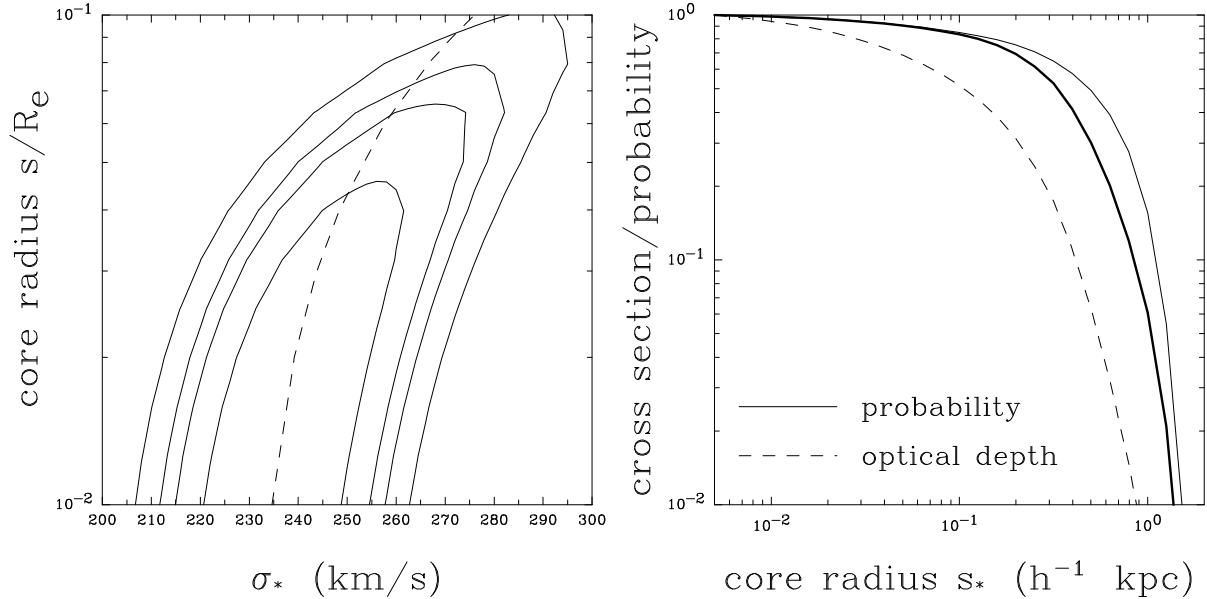

FIG. 5.–(Left) Contours of the $\chi^2$ for the dynamical fits. The light solid lines show the 68%, 90%, 95%, and 99% confidence limit changes on $\Delta\chi^2$ for the fit to the sample. The dashed line shows the expected scaling of $\sigma_{DM*}$ with $s/R_e$ if we keep the average velocity dispersion interior to $R_e$ fixed in the Hernquist/softened isothermal sphere dynamical model.

FIG. 6.–(Right) Variation in cross section (dashed line) and lensing probability (solid line) with core radius $s$ for a lens with $\sigma_{DM} = 250$ km s$^{-1}$. The values are normalized to unity at the minimum core radius. The heavy solid line shows the lensing probability excluding image systems with detectable central images using the same normalization as for the total probability. The results are given for the average over the quasar data sample including selection effects.

For a core radius of $s_* = 100 h^{-1}$ pc the fractional increase in the velocity dispersion is 7.5% or 17 km s$^{-1}$, less than the uncertainty in the value of $\sigma_{DM*}$. Nonetheless, its effects on models with a finite core radius are striking; it produces a 33% increase in the expected number of lenses if we keep the ratio of the core radius to the critical radius fixed ($s/b$ constant). This underestimates the effect of increasing $\sigma_{DM*}$ because the ratio $s/b$ also decreases, leading to a further increase in the number of lenses. As we found in the earlier dynamical models of E/S0 galaxies in isothermal mass distributions (Kochanek 1992, 1994, Breimer & Sanders 1993, Franx 1993), it is absolutely essential either to fit the dynamical variables using the observed distribution of image separations or to do the self-consistent stellar dynamical problem.

Self-consistency in the lensing calculation also requires a velocity dispersion that in-creases as the core radius becomes larger. A consistent model of the lenses must keep the average image separations fixed as the core radius increases, and the image separation is approximately twice the tangential critical radius of the lens ($\Delta\theta \simeq 2b(1-2\beta)$). If the core radius is increasing (larger $\beta$), then the only way to maintain constant average image separations is to also increase the average velocity dispersion (larger $b$). If we model this by keeping the tangential critical radius $r_+ = b(1-2\beta)$ fixed, then the lens cross section $\tau$ decreases as $\tau \propto (\beta-1/2)^2$ instead of $(\beta-1/2)^3$ for $\beta \simeq 1/2$. The condition that the average image separation must be constant significantly reduces the decline in the cross section caused by the inclusion of a core radius.

### 5.2 The Lensing Probabilities for Softened Isothermal Spheres

Self-consistent calculations of the lensing probability such as Kochanek & Blandford (1987), Kochanek (1991), Wallington & Narayan (1992), Kassiola & Kovner (1993), and (in most regimes) Maoz & Rix (1993) automatically include the effects of the core radius on the magnification bias, but most treatments of softened isothermal models examined only the effects of a core radius on lensing cross sections (e.g. Dyer 1984, Blandford & Kochanek 1987, Hinshaw & Krauss 1987, Krauss & White 1992, Fukugita & Turner 1991, Fukugita et al. 1992, Torres & Waga 1995). Core radii have a powerful effect on the cross section for multiple imaging and the cross section drops rapidly even when $\beta = s/b$ is significantly smaller than the threshold. Near the threshold of $\beta = 1/2$, the cross section decreases as $\tau \propto (1/2 - \beta)^3$. However, using the change in the optical depth grossly overestimates the effects of a finite core radius on the lensing probability in bright quasar samples. The core radius first eliminates images with low total magnifications, but the bright quasar samples are dominated by highly magnified images of fainter quasars and magnification bias significantly reduces the effects of adding a core radius on the probability.

We can understand this analytically in the Hinshaw & Krauss (1987) model. The tangential critical line of the lens is $r_+ = b(1-2\beta)^{1/2}$, the radial critical line is $r_- = b(\beta - \beta^2/2 - \beta^{3/2}(4+\beta)^{1/2})^{1/2}$ and the cross section is $\sigma_{mult} = \pi u_-^2 = \pi b^2(1 + 5\beta - \beta^2/2 - \beta^{1/2}(4+\beta)^{3/2}/2)$, where $u_-$ is the radial caustic. Inconsistent models of the effects of a core radius estimate the lensing probability by using this optical depth multiplied by the magnification bias for the singular model. When images 2 and 3 are merging on the radial caustic, image 1 is located at $r_1 = r_{out} = 2\beta u_-/r_-^2$, and the average magnification produced by the lens is $\langle M \rangle = r_{out}^2/u_-^2 = 4\beta^2/r_-^4$. When the core radius is small, the average magnification is 4, and as the core radius approaches the threshold of $\beta = 1/2 - \epsilon$ the average magnification diverges as $\langle M \rangle \propto \epsilon^{-2}$. The magnification probability distribution is approximately $P(> M) = (M_0/M)^2$ for $M > M_0$ where the minimum magnification is $M_0 = \langle M \rangle/2$ for fold caustic statistics (see Schneider, Ehlers, & Falco 1992, Blandford & Narayan 1986, Blandford & Kochanek 1987). If we assume a single power law quasar number counts distribution with $dN/dm \propto 10^{\gamma(m-m_0)}$ then the magnification bias varies with the average magnification as $B(m) \propto \langle M \rangle^{2.5\gamma}$ for $\gamma < 0.8$. As the core shrinks, the average magnification increases, which drives up the magnification bias. The lensing probability, including the change in the magnification bias, varies as $\tau B(m) \propto \epsilon^{3-5\gamma}$ not $\tau \propto \epsilon^3$. For large average magnifications the effective value of $\gamma$ is the faint slope of the quasar number counts, $\gamma \simeq \beta \simeq 0.27 \pm 0.07$ (see

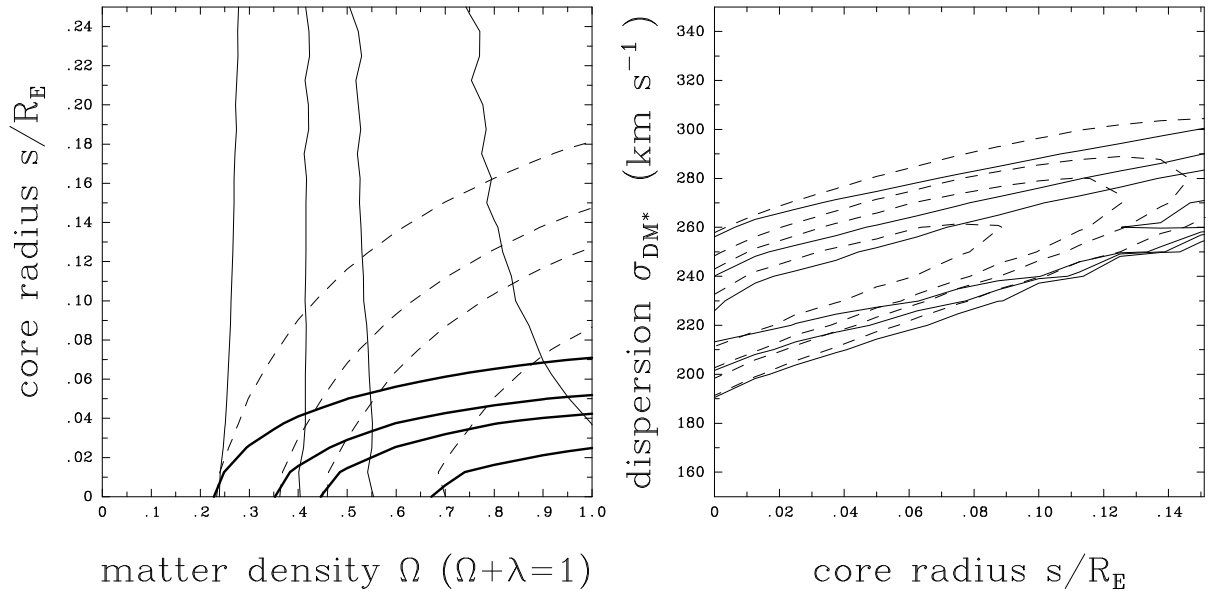

FIG. 7.–Cosmological effects of finite core radii. The left panel shows contours as a function of the ratio of the core radius to the effective radius and the cosmological model. The light solid lines are the constraints from lensing alone, the dashed lines adds the velocity dispersion prior as a function of core radius estimated in §5.1, and the heavy solid lines add the goodness of fit to the dynamical data. The right panel shows dependence of $\sigma_{DM*}$ on the core radius in an $\Omega_0 = 1$ cosmology. The light contours use only the lens data, and the dashed lines include the prior probability distribution for $\sigma_{DM*}$ estimated from dynamical models. The contours are drawn at the 68%, 90%, 95%, and 99% confidence levels for one parameter.

§2.5), and $\tau B(m) \propto \epsilon^{1.65\pm0.35}$. For bright quasars the increase in the bias is greater because of the steeper number counts slope, and the effects of the core radius are still smaller.

To emphasize this point, Figure 6 shows the relative variation of the cross section and the true lensing probability including magnification bias for a lens with $\sigma_{DM} = 250$ km s$^{-1}$ as a function of the core radius averaged over the full quasar data sample. For a core radius of $s = 100h^{-1}$ pc using the cross section instead of the true probability underestimates the lensing probability by about 40%. This comparison still overestimates the effect of a finite core radius because we did not include the dependence of $\sigma_{DM}$ on $s$ from §5.1. Near $s = 100h^{-1}$ pc we estimated that $\sigma_{DM}$ should increase by 7%, and the net lensing probability with a $100h^{-1}$ pc core radius may be higher than for a singular model.

### 5.3  The Cosmological Effects of Softened Isothermal Spheres

We assume that the core radii of galaxies are proportional to their effective radii $s = s_*(L/L_*)^{1.2}$, and the models are characterized by a fixed ratio of $s/R_e$ (as in Fukugita & Turner 1991). To simplify the calculations, we set the "Faber-Jackson" exponent to be $\gamma = 4$, so the core radius varies with the velocity dispersion as $s = s_*(\sigma_{DM}/\sigma_{DM*})^{4.8} = s_*(L/L_*)^{1.2}$. We know from §3.1 that the effects of the uncertainties in $\gamma$ are minimal. If $s/s_* =$ $(\sigma_{DM}/\sigma_{DM*})^{2+u}$ then the strength of the lens varies as $\beta = s/b = (s_*/b_*)(\sigma_{DM}/\sigma_{DM*})^u$, and when $u > 0$ the core radius suppresses lensing by large ($L > L_*$) galaxies. As we vary the velocity dispersion, the maximum tangential critical radius is

$$\max(r_+) = b_* \beta_*^{-1/u} u^{1/2} (2+u)^{-1/2-1/u} = 0.436 b_* \beta_*^{-0.36} \tag{17}$$

when $u = 2.8$. The maximum image separation is roughly twice the maximum critical radius, so to produce images with separations of 2 arcseconds for a source at $z_s = 2$ and with $\Omega_0 = 1$ requires $\sigma_* \gtrsim 257(s/h^{-1}\text{kpc})^{0.13}$ km s$^{-1}$. Thus the scaling law for the core radius introduces a sharp threshold on the velocity dispersion as a function of the core radius.

The right panel of Figure 7 shows the best fit value of $\sigma_{DM*}$ as a function of $s_*$ in an $\Omega_0 = 1$ cosmology. As expected from §5.1, the velocity dispersion increases as the core radius increases with $\sigma_{DM*} \propto 1 + s/R_e$. This is shallower than the slope seen in the dynamical models. The left panel of Figure 7 shows the dependence of the cosmological limits on the core radius using only the lens data, the lens data combined with the prior probability distribution for $\sigma_{DM*}$ derived from the dynamical model of §5.1, and finally the lens data, the dynamical velocity dispersion prior, and the likelihood of the dynamical model. If we use only the lensing data, the cosmological limits are nearly independent of the core radius – this is a radically different picture of the effect of a core radius than that found in inconsistent calculations. There is some decrease in the expected number of lenses, and the best fit models near $\Omega_0 = 1$ prefer a finite core radius to produce a modest reduction in the expected number of lenses. Since the dynamical estimate of the velocity dispersion rises more rapidly than the lensing estimate, adding the dynamical estimate of the velocity dispersion eliminates the large core radius solutions because of the rapidly increasing number of lenses. If we weight the models by the likelihood that they fit the dynamical data, the high $\chi^2$ of models with $s/R_e \gtrsim 0.05$ seen in Figure 5 dominates the limits on the core radius.

In summary, quantitative estimates of the effects of core radius must include both the increases in the velocity dispersion and the increases in the magnification bias caused by adding a finite core radius. Most previous calculations have ignored both of these corrections, and no previous calculation has included the increase in the velocity dispersion. Figure 7 covers a much broader range of core radii than is physically reasonable for E/S0 galaxies. The reasonable range of $s_* \lesssim 200h^{-1}$ pc, or $s/R_e \lesssim 0.05$, is consistent with galaxy dynamics, lensed images separations, and the absence of central images (Wallington & Narayan 1993, Kassiola & Kovner 1993). Over this range the cosmological limits are independent of the core radius.

## 6 Spiral Galaxies

We have hitherto ignored the contribution from spiral galaxies. This is a conservative approach, since the additional cross section of the spirals will only strengthen the constraints on the cosmological constant. The opposite limit and the maximum contribution is found by modeling the spirals as singular isothermal spheres with the characteristic velocity dispersion $\sigma_{DM*} = (144 \pm 10)$ km s$^{-1}$ and Tully-Fisher (1976) exponent $\gamma = 2.6 \pm 0.2$ used by Fukugita & Turner (1991) combined with our more recent estimates of the spiral density $n_s = (0.79 \pm$ $0.26)h^3 \times 10^{-2}$ Mpc$^{-3}$ and Schechter function exponent $\alpha = -1.00 \pm 0.15$ from §2.4. Because of computational limits we fixed $\sigma_{DM*} = 144$ km s$^{-1}$ and $\gamma = 2.6$ for the spiral calculation, made the variations in the Schechter function exponent for the E/S0 and spiral classes fully correlated (same $\alpha$ for both populations), and independently varied the spiral density using a log-normal distribution. There are too many variables to independently average over the prior probability distributions for all parameters of the combined E/S0 and spiral model simultaneously. We varied $n_s$ and $n_e$ independently with the broad error bars implied by Marzke et al. (1994). We could reduce the variance by using the total density $n_s + n_e$ and the fraction of E/S0 galaxies $n_e/(n_e + n_s)$ as the independent variables since the dominant errors come from the fraction of E/S0 galaxies rather than the total density (see §2.4). At the average values of the parameters the model predicts that 7 lenses are produced by the E/S0 population and 1 lens is produced by the spiral population when $\Omega_0 = 1$. This is close to the ratio of the optical depths given the parameters of the two populations. Figure 8 shows that the cosmological limits change by only $\Delta\lambda_0 \simeq 0.05$ when we add the maximum contribution from spiral galaxies.

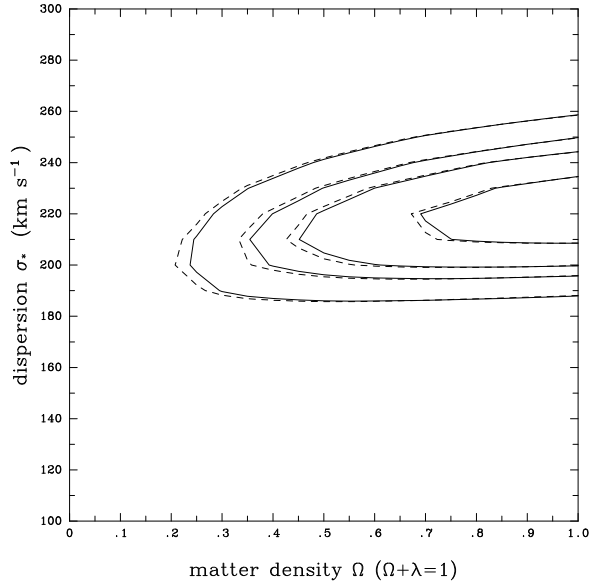

FIG. 8.–Effects of spiral galaxies. The solid (dashed) lines show the likelihood contours with (without) the maximal contribution of spiral galaxies as a function of the matter density and the velocity dispersion of the E/S0 population. The net change in the cosmological limits is about 0.05 in $\Omega_0$. Contours are shown at the 68%, 90%, 95%, and 99% confidence limits for one variable in the likelihood ratio. The likelihoods include the full uncertainties in the number counts and "Faber-Jackson" relation for the E/S0 galaxies, but only the number counts uncertainties for the spiral galaxies.

Since even the maximum contribution from the spirals does not significantly alter the cosmological conclusions, we did not pursue their effects in greater detail for all cosmological models. We did, however, consider how their contribution is modified by extinction and finite core radii. We know that spiral galaxies contain significant amounts of dust, with central extinctions of order $A_B \simeq 1$ magnitude (Peletier & Willner 1992) and scale lengths of $\sim 30$ kpc (Zaritsky 1994, Peletier et al. 1995). When we approximate the reddening in spirals by a uniform extinction of $A_B = 0.5$ magnitudes, the expected number of lenses drops to one third the number of lenses without extinction. Maoz & Rix (1993) examined spiral galaxies with core radii having the same scaling as the effective radii of E/S0 galaxies, while Krauss & White (1992) examined models with core radii independent of the luminosity. We considered the intermediate case where $s = s_*(\sigma/\sigma_*)^2$, so that $\beta = s/b$ is independent of luminosity and all spiral galaxies are equally affected by the finite core radius. We do not have to change the values of $\sigma_*$ to compensate for the core radius because the normalization is determined by the asymptotic rotation velocity of the galaxies outside the core regions. For $s_* \lesssim 70h^{-1}$ pc the expected number of lenses changes by less than 20%, but above this threshold the numbers begin to drop precipitously. The number of lenses has dropped by 50% at $s_* \simeq 200h^{-1}$ pc, and by 90% at $s_* \simeq 500h^{-1}$ pc.

Between the effects of the smaller mass compared to E/S0 galaxies, reddening, and finite core radii, the effects of the spiral galaxies on cosmological limits are negligible. The uncertainties in lens statistics must be closer to 20% than 50% before there is any need to include spiral galaxies in statistical models. The exception to this rule is for models that use large systematic errors to suppress the contribution of the E/S0 galaxies by factors of two or more; in these models the contribution of the spiral galaxies is important.

## 7  Lens Galaxy Luminosity and Extinction

Lensed quasars differ from unlensed quasars not only because they are multiply imaged but because of the luminosity and absorption properties of the lens galaxy. These effects are less important for radio selected lenses, but they can introduce large systematic errors in statistical analyses of optically selected lenses (Kochanek 1991). In bright quasar surveys ($m \lesssim 19$ B mags) we do not expect the luminosity of the lens galaxy to be a significant problem because the source quasars are almost always significantly brighter than the lens galaxy. Reddening, however, can easily reduce the magnification bias by more than a factor of two. The question of reddening is whether the reduction in the number of lenses is consistent with the properties of E/S0 galaxies locally and with other properties of the lens systems such as their separations, lens galaxy redshifts, and the similarity of the radio and optically selected samples. For simplicity we use the nearly singular isothermal models from §3.

### 7.1  The Luminosity of the Lens Galaxy

The current optical lens surveys select objects from catalogs of known quasars, so the luminosity of the lens galaxy affects the statistical model if it selectively excludes lensed quasars from catalogs. At the lens survey stage, the detectability of an extended galaxy increases the completeness of the survey because it helps to distinguish the lens from a point source. All quasar catalogs have a limit on the luminosity of the lens galaxy beyond which the object will not be recognizable as a quasar. Color and variability selected samples are more sensitive to the brightness of the lens galaxy than spectrally selected samples (this includes X-ray and radio selected samples). Where identifying the object as a quasar depends only on finding quasar emission lines, the galaxy must be significantly brighter than the quasar to conceal the strong, broad emission lines.

The luminosity of the lens is not a serious problem in the current optical lens surveys because they all selected bright quasars with $m \lesssim 18$ B mags. Figure 9 shows the approximate effects of the lens luminosity by assigning the lens galaxy the luminosity $M_{B_T} = -19.9 - 2.5\gamma \log(\sigma_{DM}/\sigma_{DM*})$ and (neglecting k-corrections and evolution for simplicity) eliminating all lenses in which the quasar magnitude was less than $\Delta m$ magnitudes brighter than the galaxy. The limit on the brightness of the lens galaxy must exceed $\Delta m = 3$ magnitudes before it introduces any serious systematic bias in the statistical analysis, and $\Delta m = 3$ exceeds any realistic observational limit even for color selected quasars. When $\lambda_0 > 0$ the lens galaxy is typically further from the observer and fainter, so for a fixed threshold $\Delta m$ you find a larger fraction of the lenses in a high $\lambda_0$ model than in a low $\lambda_0$ model.

We calculated the likelihoods including magnitude limits of $\Delta m = 0$, 1, 2, and 3 magnitudes on the luminosity of the lens galaxies. The models included the $k$-corrections from Coleman, Wu, & Weedman (1980), but no evolutionary corrections. The changes in the cosmological limits are smaller than we can resolve given the coarse gridding of the simulations (comparable to the right panel of Figure 3) so there is little point in showing the likelihood contours. The $\Omega_0 = 1$ models are the most sensitive to the selection effect; yet when $\Delta m = 2$ only a few percent of the lenses are undetectable. Note, however, that the selection effect becomes a serious concern in surveys for fainter lenses. For magnitudes between $19 \lesssim B \lesssim 20$ it is an important correction, and for fainter magnitudes ($B \gtrsim 20$) the statistical results may be controlled by the accuracy of the corrections for this systematic bias. This confirms the conclusions of Kochanek (1991).

## 7.2 Reddening

Reddening alters the detectability of a lens in three ways. The mean reddening reduces the average magnification bias (Kochanek 1991, Tomita 1995), radial gradients in the reddening generally increase flux ratios (Tomita 1995, Fukugita & Peebles 1995), and patches like molecular clouds can absorb single images. In a bright quasar sample the lensed images are found near the critical radius of the lens because the magnification and hence the magnification bias is large there. Since the images lie near the tangential critical radius, gradients in the extinction are more important for producing variations in the mean opacity with lens redshift than for increasing the flux ratios. The cross section averaged distance of the critical line from the center of the lens galaxy is $2.0h^{-1}(\sigma_{DM}/225 \text{ km s}^{-1})^2$ kpc for $\Omega_0 = 1$ and a source at $z_s = 1$. The numerical coefficient rises to $2.7h^{-1}$ kpc for $z_s = 4$. Adding a cosmological constant increases the average impact parameter, with the limiting coefficients of $3.1h^{-1}$ kpc ($6.4h^{-1}$ kpc) for $\lambda_0 = 1$ and $z_s = 1$ (4). There must be significant dust opacity on scales comparable to the effective radius $R_e$ to significantly affect lens statistics. Dust concentrated in the central regions of the galaxy will have little or no effect.

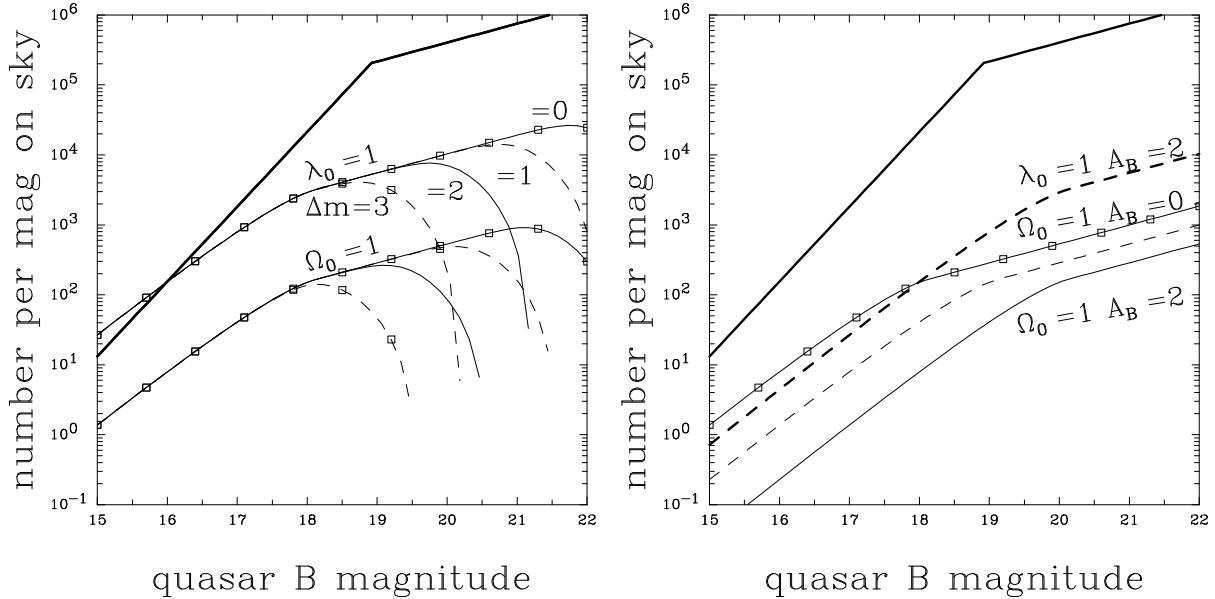

FIG. 9.–The qualitative effects of the lens galaxy luminosity (left panel) and extinction (right panel) on the expected number of lenses as a function of magnitude. The models are normalized to estimate the number of lenses per magnitude on the sky normalized by 5 quasars per square degree at 19 B mags and a fixed quasar redshift of $z_s = 2$. In both panels the heavy solid curve at the top shows the estimated number counts of unlensed quasars. In the left panel the number counts of lenses are shown for $\Omega_0 = 1$ and $\lambda_0 = 1$ with the requirement that the lens galaxy is $\Delta m = 0, 1, 2$, or 3 magnitudes fainter than the quasar. The model includes neither $k$-corrections nor evolution. In the right panel the number counts of lenses are shown for $\Omega_0 = 1$ and average extinctions of $A_B = 0$, 1 or 2 magnitudes and for $\lambda_0 = 1$ and an average extinction of 2 magnitudes. The model assumes the extinction is homogeneous and independent of redshift. The model parameters are $n_e = 0.0061 h^3$ Mpc$^{-3}$, $\alpha = -1$, $\gamma = 4$, and $\sigma_{DM*} = 225$ km s$^{-1}$. Only lenses larger than 0.5 arcsec in separation and with flux ratios smaller than 100:1 are included.

Patchy absorption is the easiest to constrain both in nearby galaxies and in gravitational lenses. For example, Goudfrooij et al. (1994) find patchy absorption at the level of $A_B \lesssim 0.2$ mags in their sample of E/S0 galaxies. We estimate that the covering fraction $f$ of opaque regions in E/S0 galaxies must be smaller than $f < 0.094$ (at $2\sigma$) from the lensed radio point sources, because whenever we see one radio image in the optical, we see all the images in the optical. We know there is differential reddening in MG0414+0534 (Lawrence et al. 1994) and 2237+030 (Nadeau et al. 1991), but the absence of detailed spectral and photometric data on most lenses makes it impossible to do a more detailed study of the extinction variations in known lenses. Direct measurement of the mean absorption is difficult to obtain. In spiral galaxies the central reddening in the B band is $A_B \simeq 1.0 \pm 0.4$ (Peletier & Willner 1992). The distribution is consistent with $A_B \propto \exp(-r/r_0)$ where $r_0 \sim 30$ kpc both by direct observations (Zaritsky 1994, Peletier et al. 1995), indirect estimates from the quasar population (Heisler & Ostriker 1988) and damped Ly$\alpha$ absorbers (e.g. Lanzetta, Wolfe, & Turnshek 1995). The absorption in E/S0 galaxies is neither as large nor as well studied. Goudfrooij et al. (1994ab, 1995) suggest that there is a diffuse component with $0.1 \lesssim A_B \lesssim 0.9$, but it is unclear whether this an average for the central regions or over many effective radii. Lauer (1988) found no signs of extinction on scales of kpc (limit $A_B \lesssim 0.04$ mag) by modeling the photometry of overlapping cluster ellipticals. The red spectra of MG0414+0534 can be produced by 3-6 magnitudes of extinction in the lens galaxy, although the uniformity of the extinction for the four images is then remarkable. The red color of some radio lenses compared to normal quasars is ambiguous because Webster et al. (1995) find that unlensed radio-selected quasars show a broad, flat color distribution with $2 \lesssim B - K \lesssim 10$ (almost as red as MG 0414+0534), and O'Dea et al. (1994) find that radio galaxies can also show very red colors. Other radio selected lenses show no signs of absorption. In 0957+561, the radio, B, and R band flux ratios are all identical even though the A image is $17.3h^{-1}$ kpc from the galaxy and the B image is only $2.6h^{-1}$ kpc from the center. This limits the differential reddening to $\Delta A_B \lesssim 0.1$ mags.

As a simple model we gave the E/S0 galaxies the extinction profile $A_B = A_0(1 - r/r_0)$ for $r < r_0$ and zero for $r > r_0$, where $A_0$ is the central value and $r_0$ is a cutoff in the dust distribution. Assuming Galactic properties for the dust and gas, a dust to gas mass ratio of $\xi_2 = 10^2 M_{dust}/M_{gas}$, and an HI column density to absorption relation of $N[HI] = 1.5 \times 10^{21} A_B$ cm$^{-2}$ mag$^{-1}$, the total HI and dust masses are

$$M_{gas} = 1.3 \times 10^7 \xi_2^{-1} A_0 \left[\frac{r_0}{\text{kpc}}\right]^2 M_\odot \quad \text{and} \quad M_{dust} = 1.3 \times 10^5 A_0 \left[\frac{r_0}{\text{kpc}}\right]^2 M_\odot \qquad (18)$$

(eg. Jura et al. 1987, Goudfrooij et al. 1994ab, 1995). For a galaxy at distance $D$ with an average dust temperature of $T = 20$ K, the expected 100 $\mu$m infrared (Jura 1986, Jura et al. 1987) and HI radio fluxes (Knapp, Turner, & Cunniffe 1985; Wardle & Knapp 1986) are

$$F[100\mu\text{m}] = 0.8 A_0 \left[\frac{r_0}{\text{kpc}}\right]^2 \left[\frac{10\text{Mpc}}{D}\right]^2 \text{Jy} \quad \text{and} \quad F[HI] = 0.6 \xi_2^{-1} A_0 \left[\frac{r_0}{\text{kpc}}\right]^2 \left[\frac{10\text{Mpc}}{D}\right]^2 \text{Jy km s}^{-1}.$$
(19)

Typical E/S0 galaxies have 100$\mu$m fluxes (Jura 1986, Jura et al. 1987) and HI fluxes (Knapp et al. 1985, Wardle & Knapp 1986) comparable to these values, implying that the absorption in E/S0 galaxies locally must satisfy the integral condition that $A_0(r_0/\text{kpc})^2 \lesssim 1$. The observation of singular cores in most E/S0 galaxies (Tremaine et al. 1994) requires that $A_0 \lesssim 1$, or we would see an apparent core radius at the dust photosphere. Figure 10 shows contours of the dust mass as a function of the central extinction and the scale length, and the local evidence suggests that E/S0 galaxies occupy the lower left corner.

To include extinction in the lens calculation we should include the wavelength dependence of the dust opacity because the redshifted wavelength of an observed B photon was further in the UV when it passed through the lens galaxy. Following Heisler & Ostriker (1988) and Tomita (1995) we used the Seaton (1979) model for the UV extinction curve as a function of wavelength. Converted into a dependence on redshift normalized by the extinction in the B band at zero redshift, $A_B(0)$,

$$A_B(z) = A_B(0) \begin{cases} 1 + z & z < 0.12 \\ g\left[3.94 + 2.38z + \frac{1.01}{(2.27z-2.33)^2+0.280}\right] & 0.12 < z < 0.61 \\ g\left[4.22 + 1.93z + \frac{1.01}{(2.27z-2.33)^2+0.280}\right] & 0.61 < z < 2.14 \\ g\left[10.44 - 4.20z + 1.53z^2\right] & 2.14 < z < 3.4 \\ 1 - 3.76g + z & z > 3.4 \end{cases} \qquad (20)$$

where $g = \lambda_B/3.2\lambda_V$.[6] For a spectral index of $\alpha = -0.5$ the effective wavelengths are $\lambda_B = 4403$ Å and $\lambda_V = 5513$ Å. The fitting formula is complicated because it models the 2200 Å absorption bump seen in our galaxy, with the peak appearing near $z = 1$. In this model the optical depths at $z = 0.5, 1.0,$ and $1.5$ are $1.4, 2.4,$ and $2.0$ times larger than at $z = 0$.

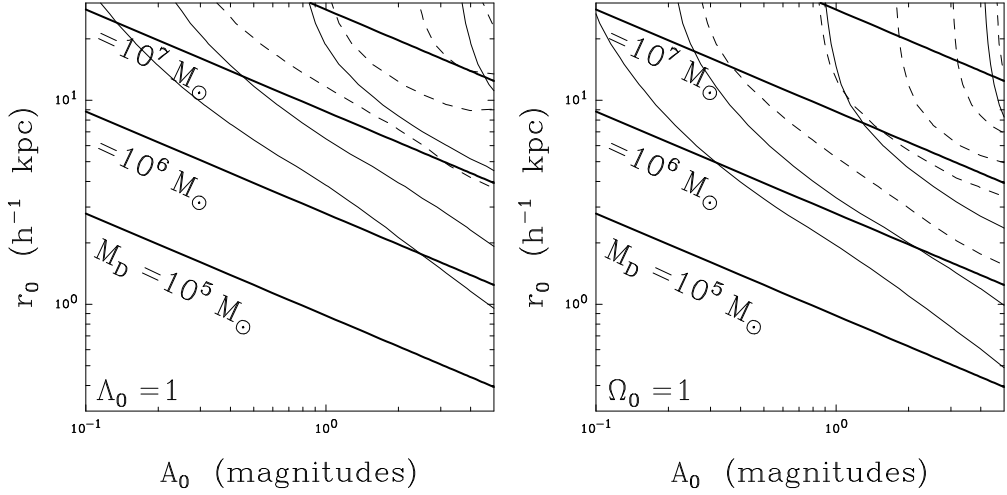

Fig. 10.—Lens probability and average image separation for $\lambda_0 = 1$ and $\Omega_0 = 1$ for the extinction profile $A_B = A_0(1 - r/r_0)$ and a 18 B magnitude source at redshift $z_s = 2$ using the standard E/S0 lens model. The values of $r_0$ and $A_0$ are for an $L_*$ galaxy. The edge radius scales as $r_0 \propto (L/L_*)^{1.2}$ and the total dust mass $\propto A_0 r_0^2$ increases as $L/L_*$. The heavy solid lines show the lines of total dust mass equal to $10^5 M_\odot$, $10^6 M_\odot$, $10^7 M_\odot$ and $10^8 M_\odot$, the light solid lines are the contours where the fraction of the lenses found drops to 90%, 80%, 50%, and 10% of the number expected without extinction, and the light dashed lines are the contours where the average separation rises to 110%, 120%, 130%, 140% and 150% of the average separation without extinction.

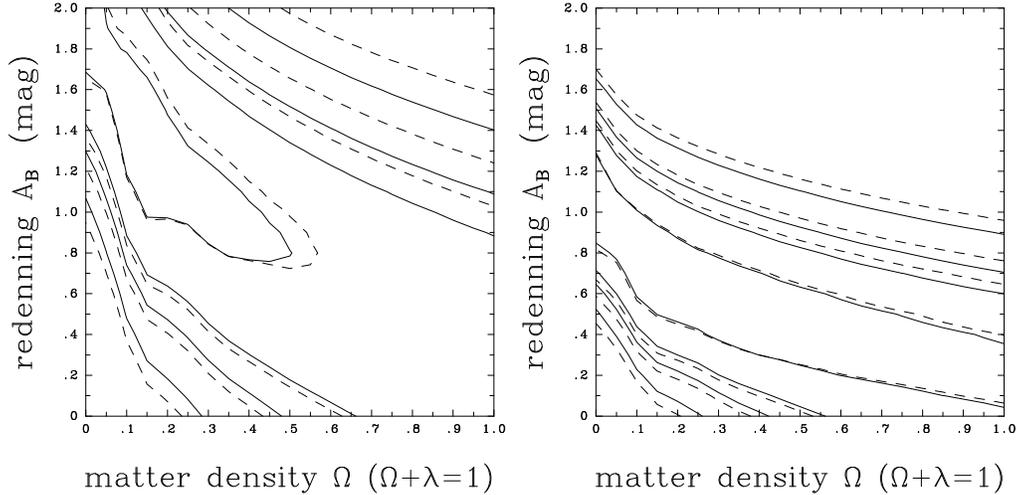

Fig. 11.—Likelihood contours for uniform reddening models. The left panel assumes a uniform extinction of $A_B$ magnitudes independent of redshift, while the right panel assumes a uniform extinction of $A_B$ magnitudes at $z = 0$ and the redshift/wavelength dependence of equation (20). The contours show the 68%, 90% confidence, 95%, and 99% confidence limits on one parameter.

Let $m_s$ be the magnitude of the source, then (for two images) the magnitude of the lensed images are $m_1 = m_s - 2.5 \log M_1 + A_1$ and $m_2 = m_s - 2.5 \log M_2 + A_2$ where $M_1$ and $M_2$ are the image magnifications with $M_2 < M_1$. We know that the total observed magnitude of the quasar is $m$ and the flux ratio of the images is $m_2 - m_1 = 2.5 \log M_1/M_2 + A_2 - A_1$. The two modifications to the lens probability calculation from reddening are to change the argument of the quasar number counts ratio in equations (4) and (5) from $m + 2.5 \log M$ to $m_s$ where $m_s$ is the source magnitude given the magnifications of the images and the extinction,

$$\frac{dN}{dm}(m + 2.5 \log M) \left[\frac{dN}{dm}(m)\right]^{-1} \rightarrow \frac{dN}{dm}(m_s) \left[\frac{dN}{dm}(m)\right]^{-1}, \qquad (21)$$

and to change the selection effects to use the reddened image fluxes.

The simplest model problem is to assume a uniform extinction of $A_B$ at all redshifts and see how it affects the magnitude distribution of lensed quasars. Figure 9 shows the qualitative effects of extinction on the numbers and magnitudes of lensed quasars. An average reddening of 2 magnitudes is needed to reduce the expected number of lenses for $\lambda_0 = 1$ cosmologies to the number expected for $\Omega_0 = 1$. The qualitative sign of extinction in the lens population is a shift in the break of the lensed quasar number counts to fainter magnitudes. This effect cannot be detected in a bright lens survey ($B \lesssim 19$ mags) because we do not survey beyond the break magnitude, but a fainter lens survey will see significantly more faint lenses in a reddened model with a cosmological constant than in a unreddened model with $\Omega_0 = 1$. In our example, the difference is a factor of 5 at 20 B mags.

Variations in the extinction with redshift or radial gradients of the extinction are more easily detected than uniform extinction because they modify the distributions of image separations or lens galaxy redshifts. For example, models in which E/S0 galaxies become nearly opaque at higher redshifts (e.g. Fukugita & Peebles 1995) are identical in their lensing effects to the galaxy formation models treated by Mao (1991) and Mao & Kochanek (1994), allowing them to be ruled out. As a model of the effects of the spatial distribution of dust we set the extinction to $A_B = A_0(1 - r/r_0)$ for $r < r_0$ where the extinction scale length varies with the effective radius $r_0 \propto (L/L_*)^{1.2}$, and the total dust mass is proportional to the luminosity $A_0 r_0^2 \propto L$. We included the redshift/wavelength dependence of the extinction from eqn. (20), and Figure 10 shows the effects of dust on the lensing probability and image separations for a source with $z_s = 2$ and $m = 18$ B mags (the typical survey quasar). A cosmological distribution of lenses with the local dust content of E/S0 galaxies ($M_{dust} \lesssim 10^5 M_\odot$ and $A_0 < 1$) changes the expected number of lenses by less than 10%. Reducing the number of lenses by a factor of two requires $M_{dust} \gtrsim 10^7 M_\odot$, but models at this lower limit alter the average image separations by 10-20% and require very high central opacities. It is impossible to reduce the expected number of lenses by a factor of ten without exceeding limits on the total amount of metals in the lens galaxy or altering the average image separations by an easily detectable amount.

---

[6]In Heisler & Ostriker (1988) the redshift limit for the first region is incorrectly given as $z < 0.19$, leading to a discontinuous function.

Because we cannot explore this parameter space fully, we confined our full statistical calculation to the effects of uniform extinction. We calculated models both with and without the redshift/wavelength dependence of the extinction curve. Remember that these cases are the hardest to limit using only the lens data, and they correspond to enormous total dust masses – *while they lead to the greatest cosmological uncertainty, they are also physically unrealistic*. We include these two cases only to illustrate the limits of the extinction problem. Figure 11 shows the likelihood contours as a function of the extinction and the cosmological model. As expected, there is a degeneracy between the two parameters, since with only five bright lenses the magnitude and redshift dependence of the lens probability are too weakly measured to constrain the model. The degeneracy is stronger when we neglect the redshift dependence of the extinction because it minimizes the side effects of the absorption on the image separations. The increase in the opacity with redshift acts like models of galaxy evolution where the number of lenses declines with redshift, and these models drive up the average lens separation allowing them to be constrained from the lens data (Mao & Kochanek 1994). If we neglect the dependence of the extinction on redshift, we underestimate the amount of extinction required to change the cosmological model by approximately a factor of two.

Both of these models overestimate the extent of the degeneracy. The $\lambda_0 = 1$, high $A_B$ models are as acceptable as the $\Omega_0 = 1$, $A_B = 0$ models not just because they both produce the same number of lenses, but because we put no limits on the reddening in the observed lenses. Two magnitudes of extinction in any of the known optical lenses would be easily observed, so we know that a significant fraction of the lens population at intermediate redshifts is transparent. Moreover, there is no systematic pattern of the radio lenses having either higher lens galaxy redshifts or fewer measurable lens redshifts than the optical lenses as is required to make the optical and radio selected samples consistent with large amounts of extinction and a high cosmological constant. The most powerful constraint on dust in E/S0 galaxies is probably that quasars with damped Ly$\alpha$ absorption systems are not lenses. Bartelmann & Loeb (1995) show that there is already a high probability that high column density damped Ly$\alpha$ systems are lensed by spiral galaxies. If the E/S0 galaxies contained comparable amounts of gas at $z \sim 1$, the probability that the damped Ly$\alpha$ systems are lensed would increase dramatically.

## 8   THE LENS REDSHIFT TEST

The previous calculations did not include information about the relative redshifts of the lens galaxy and the source. Kochanek (1992) pointed out that all the known lens redshifts were significantly lower than expected for models with a large cosmological constant. The analysis neglected both parameter uncertainties and the detectability of the lens galaxy. King (1994) and Helbig & Kayser (1995) point out that the selection effects can be important.

Let $p_i(z)$ be the normalized likelihood that the known lens with separation $\Delta\theta_i$ and source redshift $z_{si}$ has a lens at redshift $z$ with $0 < z < z_{si}$, and let the magnitude of the lens at that redshift be $m_i(z)$. Let $P_{Mi}(m_i)$ be the probability that the lens redshift can be measured for a lens of magnitude $m$ in lens system $i$. Then the probability that we can measure the lens redshift for system $i$ is $P_{Di} = \int dz P_{Mi}(m_i)p_i(z)$ and the probability that we cannot measure the lens redshift is $P_{Ui} = 1 - P_{Di}$. We can also include the probability that the system is found as a lens by adding an additional term to exclude redshifts where the lens galaxy is brighter than the quasar, but this is such a negligible part of the total probability in the current sample of lenses that we do not include it (see §7.1).

Suppose we have $N$ lenses with known source redshifts, and that $N_D$ have measured lens redshifts and $N_U$ do not. The Kochanek (1992) approach defined the likelihood function by $L_{old} = \Pi_{j=1}^{N_D} p_j(z_j)$. Helbig & Kayser (1995) point out that this does not account for where the redshifts are measurable, and they advocate only looking at the redshift probability distributions over the region in which they are measurable, defining the likelihood function to be $L_{HK} = \Pi_{j=1}^{N_D} P_{Mj}(m)p_j(z_j)/P_{Dj}$. The insensitivity found by Helbig & Kayser (1995) is, however, an artifact of the $L_{HK}$ statistic because it does not account for the fraction of the lenses with measurable redshifts. We should be unable to measure the lens redshifts in a much higher fraction of the lenses if $\lambda_0 \sim 1$ than if $\lambda_0 \sim 0$. Hence when we include the selection effects we should also add the probability of failing to measure the redshifts in the systems where we cannot find the lens. We define the likelihood

$$L_{new} = \Pi_{j=1}^{N_D} p_j(z_j) \Pi_{i=1}^{N_U} P_{Ui} \qquad (22)$$

which combines the probability of the $N_D$ measured redshifts with the probability that the $N_U$ unmeasured redshifts were unmeasurable. By including the information on the fraction of the redshifts that are measurable, we restore the sensitivity to the cosmological model.

We retreat to the SIS model for this analysis. The small core radii permitted by the large statistical calculations of §5 will only lead to slight modifications of the tails of the probability distribution. The normalized differential probability that a lens has image separation $\theta$ and redshift $z$ given a single Schechter (1976) function distribution of galaxies and a "Faber-Jackson" relation is (Kochanek 1992)

$$p_i(z) = \frac{D_{OL}^2}{(1 + \Omega_k D_{OL}^2)^{1/2}} \frac{dD_{OL}}{dz} x^{1+\alpha} \exp(-x) \left[ \int_0^{D_{OS}} \frac{D_{OL}^2 dD_{OL}}{(1 + \Omega_k D_{OL}^2)^{1/2}} x^{1+\alpha} \exp(-x) \right]^{-1} \qquad (23)$$

where $x = L/L_* = (\Delta\theta_i D_{OS}/2b_* D_{LS})^{\gamma/2}$ is the galaxy luminosity required to produce image separation $\Delta\theta_i$, $b_* = 4\pi(\sigma_{DM*}/c)^2$, and $D_{OL}$, $D_{LS}$ and $D_{OS}$ are the usual proper motion distances defined in §2. We average over the uncertainties in $\alpha$ and $\gamma$ (as in §3), and we include the likelihood of the observed lens separations by adding the term $\Pi_{i=1}^{N_L+N_R} p_k(\Delta\theta_k)/p_k$ (see §2.3) for the probability that the $N_L$ optical and $N_R$ radio lenses have separations $\Delta\theta_k$. This term is also independent of cosmology (for flat universes and the SIS model), the galaxy number density, and the average reddening. It serves only to constrain the value of $\sigma_{DM*}$ without needing prior dynamical information. We also compute the likelihood adding the standard prior probability for $\sigma_{DM*}$.

We use the galaxy luminosity model from Kochanek (1992) based on k-corrections from the spectral energy distributions of Coleman, Wu, & Weedman (1980) to estimate the R magnitude of the lens galaxy as a function of redshift $m_i(z)$ (see §2.4). For each lens we assign a detection threshold $m_{i0}$ (see Table 1). The probability of measuring the lens redshift is unity if $m < m_{i0}$ and has a Gaussian cutoff with a width of one magnitude $\exp(-(m-m_{i0})^2/2)$ if $m > m_{i0}$. The width of the cutoff will compensate for magnitude uncertainties and the neglect of galaxy evolution. The correct way to determine the cutoff magnitude is to take the best spectrum for each quasar and estimate the threshold at which the lens galaxy would be detected, but it is impossible to obtain the raw spectra of many of the lenses. For the systems where the lens galaxy is seen (PG 1115+080, 0142−100, MG 1654+1346, and CLASS 1608+656) we estimated how much fainter the observed galaxy could be and still have a measurable redshift. Depending on the image geometry the limit was either one or two magnitudes fainter. For the undetected lenses 1208+1011, H 1413+117, and B 1422+231 we set the limit to be 2.5 magnitudes fainter than the quasar. For LBQS 1009−0252 and CLASS 1600+434 we used stricter limits of R=20 and R=19 magnitudes based on inspection of the existing spectra.

Figure 12 shows the likelihood as a function of $\Omega_0$ and $\sigma_{DM*}$. The $L_{new}$ likelihood is sensitive to the cosmological model. It sets an upper bound on the cosmological constant of $\lambda_0 < 0.77$ at one standard deviation ($\lambda_0 < 0.90$ at $2\sigma$) and the peak is at $\lambda_0 = 0.4$. We expect the redshift test to be less sensitive than the full statistical analysis, so these weaker limits are not surprising. Large separation lenses have the brightest lens galaxies. In these systems the lens redshifts are both known and low (0142-100, PG 1115+080, MG 1654+1346, and CLASS 1608+656). Small separation and very bright lenses have lens galaxies too faint to measure a redshift in all cosmologies (1208+1011, B 1422+231 and H 1413+117). The two intermediate separation systems (LBQS 1009−0252 and CLASS 1600+434) were recently discovered. They have not been studied in detail, but the lens galaxies can probably be detected in these systems, particularly if $\Omega_0 \sim 1$. As found by King (1994) and Helbig & Kayser (1995), the $L_{HK}$ statistic is very insensitive to the cosmological model, and it cannot be used to set any statistically significant limit on $\lambda_0$.

## 9 Conclusions

The statistics of gravitational lenses, including their numbers redshifts, magnitudes and separations are consistent with the expectations for a transparent, constant comoving density of non-evolving E/S0 galaxies in a flat $\Omega_0 = 1$ cosmology. The upper limit on the cosmological constant in a flat cosmology is $\lambda_0 \lesssim 0.65$ at 95% confidence. Open universes without a cosmological constant and $\Omega_0 \gtrsim 0.2$ have likelihoods near the 90% confidence limit. These limits for the first time include all the statistical errors in the input variables determining the number and distribution of galaxies and quasars, as well as the statistical uncertainties in the lens sample. Is there a cosmic concordance? Perhaps, but not at $\lambda_0 = 0.65$.

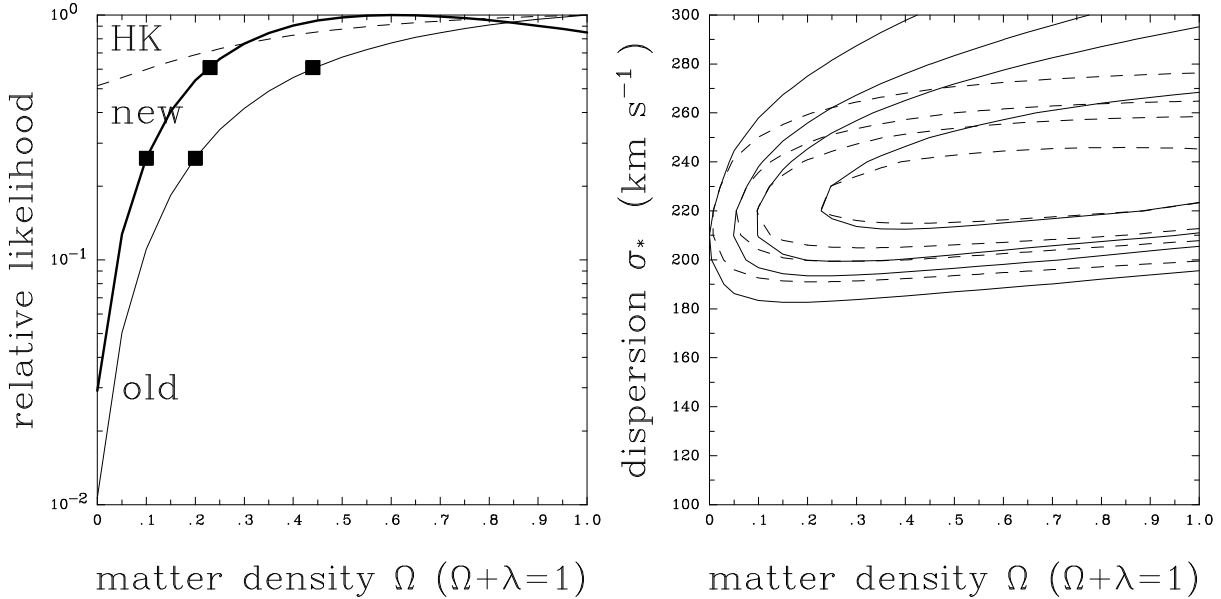

Fig. 12.—The right panel shows the likelihood contours of $L_{new}$ for the observed lens redshift sample including selection effects. Contours are shown at the 68%, 90%, 94.4%, and 99% confidence limits for one variable in the likelihood ratio. The solid lines show the limits using the lensing data alone, and the dashed lines show the limits including the prior probability distribution for $\sigma_{DM*}$. The prior probability distributions for $\alpha$ and $\gamma$ are included in both distributions. The left panel compares the cosmological limits for the same selection model using the old Kochanek (1992) statistic, the new statistic, and the HK (Helbig & Kayser 1995) statistic. The points on the curves show the one standard deviation and 90% confidence limits in the likelihood ratio. Each curve is normalized to a peak likelihood of 1.

As found by Maoz & Rix (1993) and Kochanek (1995), constant mass-to-light ratio models of galaxies modeled with de Vaucouleurs profiles are not consistent with the observed separations of the lensed images. An $L_*$ galaxy must have mass $3.1^{+2.2}_{-1.3} \times h^{-1} 10^{11} M_\odot$ or mass-to-light ratio $(M/L)_{B*} \simeq 22^{+16}_{-9} h$ at 95% confidence to fit the lens data even after allowing for all the model and cosmological uncertainties. This matches direct estimates of the mass of the lens galaxy in MG 1654+1346 (Kochanek 1995), but is inconsistent with local dynamical estimates (e.g. van der Marel 1991). The cosmological limit found using the de Vaucouleurs profile is $\lambda_0 \lesssim 0.62$ at 95% confidence, little changed from the isothermal models.

If we model the galaxies as softened isothermal spheres we find that the best fit velocity dispersion of the dark matter is $\sigma_{DM*} \simeq 220 \pm 20$ km s$^{-1}$ for an $L_*$ galaxy. Unlike the de Vaucouleurs models, we find the same estimates for $\sigma_{DM*}$ whether we fit the velocity dispersions of E/S0 galaxies (Kochanek 1993, 1994, Breimer & Sanders 1993, Franx 1993) or the image separations of gravitational lenses. The best fit models are singular ($s_* \lesssim 100h^{-1}$ pc), consistent with HST photometry of E/S0 galaxies (Tremaine et al. 1994), limits on the presence of central images in lenses (Wallington & Narayan 1993, Kasiola & Kovner 1993), models of individual lenses (eg. Kochanek 1995), and general theoretical expectations (eg. Dubinski & Carlberg 1991). Adding a core radius has no significant effects on the cosmological limits.

Earlier lens studies were concerned with limits on large cosmological constants ($\lambda_0 \sim 0.8$) where the change in the number of lenses is so great that 50% errors in the calculations are not very important. Now that we are worrying about $\lambda_0 \sim 0.5$, such inaccuracies are no longer tolerable. There are four common problems with lens calculations. (1): The $(3/2)^{1/2}$ correction factor introduced by Turner, Ostriker & Gott (1984) based on simplified models of galaxy dynamics does not exist. Models using the $(3/2)^{1/2}$ correction will overestimate the expected number of lenses by 125% and the expected separations by 50%. (2): The effects of a core radius cannot be modeled using the changes in the lensing cross section. The lensing probability is the product of the cross section and the magnification bias, and as the cross section drops the rising magnification bias prevents a dramatic reduction in the lensing probability. For a core radius of $s_* = 100h^{-1}$ pc and $\sigma_{DM} = 250$ km s$^{-1}$, using the cross section instead of the true lensing probability overestimates the drop in the probability by 40%. (3): In a consistent model, the velocity dispersion must increase when a core radius is added to the model. We find the rough scaling that $\sigma_{DM} \propto 1 + \xi(s/R_e)$ where $1 < \xi < 3$ depending on the details of the normalization, and $s/R_e$ is the ratio of the core radius to the effective radius. Because the lens probability is proportional to $\sigma_{DM}^4$, small increases in the velocity dispersion have dramatic effects on the lensing probability. The small rise in the velocity dispersion coupled with the effects of magnification bias, cause the weak dependence of the cosmological limits on the core radius. (4): The quasar number counts model of Fukugita & Turner (1991) uses the slope for the apparent magnitude number counts of all $z < 2.2$ quasars rather than the true luminosity function, causing it to underestimate the magnification bias in bright lens surveys. When we use a new number counts model derived from Hartwick & Schade (1990), we find that the Fukugita & Turner (1991) model underestimates the limits on the cosmological constant by $\Delta\lambda_0 \simeq 0.15$. We reach the same conclusion with the Boyle et al. (1988, 1990) luminosity function used by Wallington & Narayan (1993), or the Boyle et al. (1987) luminosity function used by Maoz & Rix (1993).

Spiral galaxies are not an important effect in the current lens sample. At most they improve the limits on the cosmological constant by about $\Delta\lambda_0 = 0.05$. This assumes the spiral galaxies have singular cores and no extinction. Given the typical extinction estimated by Peletier & Willner (1992) and Zaritsky (1994), their contribution drops by a factor of three. For core radii larger than $s_* \simeq 200h^{-1}$ pc their contribution drops by a factor of two. Between extinction and core radii, there is no need to include spiral galaxies in statistical models given the other sources of errors.

Astronomy is, of course, replete with examples where hidden systematic errors lead to incorrect conclusions even though all possible attention was given to the statistical uncer-tainties. We have, however, explored a wide suite of possible systematic errors. Many, such as the radial mass distribution of the lens galaxies, ellipticity, core radii, and the luminosity of the lens galaxy, can be shown to be unimportant in altering the cosmological limits. Reddening in E/S0 galaxies is a small 10% correction to lens statistics if the extinction is limited by local observations of E/S0 galaxies to have $M_{dust} \lesssim 10^5 M_\odot$ and central opacities $A_0 \lesssim 1$.

The most important remaining class of systematic errors involve evolution, in the number, mass, or extinction of the lens galaxies. Some of these issues have already been explored (see Mao 1991, Mao & Kochanek 1994, Rix et al. 1994, Fukugita & Peebles 1995), although the proliferation of parameters makes a thorough investigation difficult. While it is easy to invent a systematic effect that can lower the expected number of lenses and permit a cosmological constant, it is hard to do so without altering the observed lens separations, redshifts, magnitudes, and the similarities of the radio and optically selected lens samples. A serious suggestion for a systematic error must demonstrate consistency with all the lens data.

As a rule, evolution models that alter the number of lenses by a factor of two will inevitably alter some other observable property of the lenses. For example, galaxy formation or an abrupt rise in the opacity of the E/S0 galaxies can reduce the number of lenses only at the price of increasing their average separations. This allowed Mao & Kochanek (1994) to rule out such evolutionary models as a means of allowing $\lambda_0 \simeq 0.8$. When using evolution to avoid the limits on the cosmological constant, we should keep in mind that one effect of a large cosmological constant is to make low redshift evolution less plausible (see Carroll, Press, & Turner 1992). Moreover, physically plausible merger models tend to alter the distribution of image separations but not the expected number of lenses. There is also increasing evidence that the E/S0 population shows little evolution even at $z \sim 1$ (Lilly et al. 1995).

The most pressing needs for new observational data (aside from more lenses, and ignoring the difficulty of the observations) are good spectral studies of known lenses, better galaxy number counts, and large redshift surveys of the sources in the radio surveys. Good spectra or multi-color photometry of the already known lenses, similar to the efforts for MG0414+0134 (Lawrence et al. 1994), are the simplest way to control the effects of extinction. Better spectra are also needed to find the lens and source redshifts, or to set good magnitude limits on the presence of a lens galaxy. The number of question marks remaining in Table 1 is disheartening. The most important statistical uncertainty in the models is the number density of E/S0 galaxies, and the lens calculations should really be done using the full error correlations between the parameters of the galaxy number counts models ($\alpha$, $n_*$, and $L_*$). Finally, the differential statistics of the radio and optical surveys will be a powerful constraint on systematic errors. Yet while the radio surveys are far more productive than the optical surveys at finding lenses, the utility of the data is limited by the missing or poor characterization of the source population compared to quasars.

Acknowledgements: I would like to thank M. Bartelmann, E. Falco, P. Schechter, and P. Thaddeus for many discussions. M. Schwartz developed an early version of the algorithms to include parameter uncertainties for his Junior Thesis. This research was funded in part